\documentclass[12pt]{IEEEtran}
\normalsize

\onecolumn
\pdfoutput=1

\usepackage[utf8]{inputenc}
\usepackage[T1]{fontenc}
\usepackage{graphicx} 
\usepackage{amsmath,times}
\usepackage{epstopdf}
\usepackage{cite}
\usepackage{amstext,amssymb,latexsym}
\usepackage{setspace}
\usepackage{stackengine} 
\usepackage{relsize} 
\usepackage{cuted,mathtools} 
\usepackage{color,soul,xcolor} 
\usepackage[font=small]{caption} 

\usepackage{amsfonts, amsmath, amsthm, amssymb} 
\usepackage{multicol,lipsum}
\usepackage{textcomp} 
\usepackage{gensymb} 
\usepackage{array,tabu,makecell} 
\usepackage{boldline,multirow} 
\usepackage{bm} 
\usepackage{subcaption} 
\usepackage{booktabs} 
 
\DeclarePairedDelimiterX\set[1]\{\}{\nonscript\,#1\nonscript\,}
\DeclarePairedDelimiter\abs{\lvert}{\rvert}

\newcommand{\LRT}[2]{%
  \mathrel{\mathop\gtrless\limits^{#1}_{#2}}
}

\DeclareMathOperator*{\conj}{conj}
\DeclareMathOperator*{\eigs}{eigs}

\allowdisplaybreaks

\doublespace

\onecolumn

\newcommand{\subparagraph}{}
\usepackage[compact]{titlesec}
\titlespacing{\section}{0pt}{1ex}{1ex}
\titlespacing{\subsection}{0pt}{1ex}{1ex}
\titlespacing{\subsubsection}{0pt}{1ex}{1ex}
\begin{document}
%
\title{An Efficient Beam and Channel Acquisition via Sparsity Map and Joint Angle-Delay Power Profile Estimation for Wideband Massive MIMO Systems}

\author{Ali~O.~Kalayci,~and~Gokhan~M.~Guvensen,~\IEEEmembership{Member,~IEEE}
\thanks{The authors are with the Dept. of Electrical and Electronics Engineering, Middle East Technical University, Ankara, TURKEY (e-mail: ali.kalayci@metu.edu.tr and guvensen@metu.edu.tr)}
}

%

\maketitle

\begin{abstract}
In this paper, an efficient beam and channel acquisition scheme together with joint angle-delay power profile (JADPP), namely \textit{scatter map}, construction are proposed for single-carrier (SC) mm-wave wideband sparse massive multiple-input multiple-output (MIMO) channels when hybrid beamforming architecture is utilized. We consider two different modes of operation, namely \textit{slow-time beam acquisition} and \textit{fast-time instantaneous channel estimation}, for training stage of time division duplex (TDD) based systems. In the first mode, where pre-structured hybrid beams are formed to scan intended angular sectors, the joint angle-delay sparsity map together with power intensities of each user channels are obtained by using a novel \textit{constant false alarm rate (CFAR) thresholding} algorithm inspired from \textit{adaptive radar detection} theory. The proposed thresholding algorithm employs a spatio-temporal adaptive matched filter (AMF) type estimator, taking the strong interference due to simultaneously active multipath components (MPCs) of different user channels into account, in order to estimate JADPP of each user. After applying the proposed thresholding algorithm on the estimated power profile, the angle-delay sparsity map of the massive MIMO channel is constructed, based on which the channel covariance matrices (CCMs) are formed with significantly reduced amount of training snapshots. Then, by using the estimated CCMs, the analog beamformer is reconstructed by means of a virtual sectorization (user-grouping via second-order channel statistics) while taking the inter-group and inter-symbol interference (ISI) into account. Finally, for the second mode of operation, two novel reduced-rank instantaneous channel estimators, operating in a proper beamspace formed by the hybrid structure, are proposed. The proposed beam and channel acquisition techniques attain the channel estimation accuracy of minimum mean square error (MMSE) filter with true knowledge of CCMs. At the same time, they allow non-orthogonal pilot sequences among different users while reducing the training overhead (which is basically constant with the number of active users in the system) considerably.

\end{abstract}


\IEEEpeerreviewmaketitle

\section{Introduction}

Multi-user massive multiple-input multiple-output (MIMO) cellular system is expected to be one of the key technologies for $5$G beyond \cite{andrews14}, since it provides large gains in spectral and energy efficiency, high spatial resolution, and allows simple transceiver design \cite{marzetta10}, \cite{larsson14}, \cite{rusek13}. To capitalize aforementioned gains, instantaneous channel state information (CSI) is requisite for multi-user precoding at downlink or multi-user decoding at uplink in a massive MIMO system for both time division duplex (TDD) and frequency division duplex (FDD) modes \cite{guvensen16_2}, \cite{noh14}. However, the CSI acquisition is recognized as a very challenging task for massive MIMO systems due to the high dimensionality of channel matrices, pilot contamination, training overhead, computational complexity and so on \cite{marzetta10}. In FDD mode, CSI is typically obtained through explicit downlink training and uplink limited feedback. However, as the number of antenna elements at base station (BS) increases, the traditional downlink channel estimation strategy for FDD systems becomes infeasible \cite{swindlehurst14}. Contrary to FDD mode, the acquisition of CSI via uplink training in TDD mode is more practical \cite{ashikhmin11}.

\subsection{Related Work}
There are various channel estimation algorithms proposed in the literature. Among them, channel covariance matrix (CCM) based methods offer additional statistical knowledge about the channel parameters, and thus achieve much better estimation accuracy when compared to the compressive sensing based approaches, which try to recover CSI from fewer sub-Nyquist sampling points \cite{rao14}, \cite{gao15}, or the angle-space methods which exploit spatial basis expansion models \cite{Xie16}, \cite{Xie17}. To embrace their benefits, however, CCMs need to be acquired first. The acquisition of CCM in full dimension still constitutes a bottleneck for the performance of massive MIMO systems even for TDD mode due to the computational burden of processing large dimensional signals and significant training overhead. There are different approaches to estimate CCMs such as using temporal averaging of received signal snapshots in full dimension \cite{Brennan74}, \cite{Li03}, employing compressive sensing algorithms \cite{romero16}, \cite{park18}, and applying approximate maximum likelihood (AML) technique formulated as a semi-definite program in low dimensional subspace \cite{Haghighatshoar18}. In \cite{Xie18}, CCMs are constructed by exploiting power angular spectrum and angle parameters of channels. However, the channel estimation accuracy is still far from that of the minimum mean square error (MMSE) estimator with true CCM knowledge. Furthermore, the work in \cite{Xie18} takes only the angular sparsity of the wideband channel into account while overlooking the joint angle-delay sparsity of the mm-wave channels as did by many researchers in the literature so far. In contrast, next generation wireless systems will inevitably be broadband in mm-wave \cite{ghosh14}, \cite{swindlehurst14_2} due to the much higher throughput requirements, thus leading the channel to be sparse both in angle and delay domain.
 
Most of the existing researches for massive MIMO systems adopt flat-fading channel model by considering the use of orthogonal frequency division multiplexing (OFDM) \cite{swindlehurst14}. However, due to the drawbacks of OFDM transmission \big(e.g., high peak-to-average-power ratio (PAPR)\big), the use of single-carrier (SC) in massive MIMO systems employing mm-wave bands, exhibiting sparsity both in angle and delay plane, was considered in \cite{Larsson2012}, \cite{caire19}, \cite{Guvensen16}, \cite{kurt19}. In these studies, the mitigation of inter-symbol interference (ISI) via reduced complexity beamspace processing (rather than temporal processing) motivates the use of SC in spatially correlated wideband massive MIMO systems \cite{caire19}, \cite{Guvensen16}, \cite{kurt19}.

Besides CSI estimation problem, one can also exploit CCMs in order to reduce the effective channel dimension so that the computational burden in massive MIMO systems is highly relieved. To facilitate this approach, two-stage beamforming concept under the name of Joint Spatial Division and Multiplexing (JSDM) has been proposed \cite{adhikary13}, \cite{nam14}, where users with approximately same channel covariance eigenspaces are partitioned into multiple groups. Then, a statistical analog beamformer can be constructed only from the long-term parameters, basically CCMs, (instead of using instantaneous CSI) in order to distinguish intra-group signals while suppressing inter-group interference. Furthermore, the training overhead, necessary to learn the effective channel of each user, is reduced considerably \cite{kurt19}. Here, the JSDM framework motivates the use of hybrid beamforming architectures \cite{Molisch17}, \cite{Li18} instead of using fully digital precoding/decoding in mm-wave, where efficient reconfigurable radio frequency (RF) architectures can be implemented at competitive cost, size, and energy.

\subsection{Contributions}
In the light of above discussion, we are seeking for an effective CCM construction technique, yielding high instantaneous CSI accuracy, by exploiting only slowly varying parameters such as angles of arrivals (AoAs), temporal delays, and average power of the multipath components (MPCs). In this paper, efficient algorithms are proposed to estimate the joint angle-delay sparsity map and power profile of SC wideband massive MIMO channel to construct CCMs, based on which an adaptive beam and instantaneous channel acquisition is carried out for JSDM architecture in mm-wave bands. Throughout the paper, we conceive two modes of operation for beam and channel acquisition, namely \textit{slow-time beam acquisition} and \textit{fast-time instantaneous channel estimation}, in training stage of TDD based massive MIMO system utilizing hybrid beamforming structure. In the first mode, initially, pre-structured hybrid search beams are utilized to scan the intended angular sector of interest. Then, an adaptive spatio-temporal matched filter (AMF), taking the simultaneously active interfering users into account, is designed to estimate power intensities of active MPCs for each user channel, which is a kind of scatter map on joint angle-delay plane. Following the joint angle-delay power profile (JADPP) estimator, a novel constant false alarm rate (CFAR) algorithm, inspired from \textit{adaptive radar detection} theory, is applied onto the estimated JADPPs in order to extract joint angle-delay sparsity map, showing the spatio-temporal locations (cells) where the power of each MPC is concentrated. After obtaining the sparsity map and power profiles, the parametric construction of CCMs for each active MPC is realized in full dimension. Here, different from the existing literature \cite{romero16}, \cite{park18}, heavily based on compressive sensing tools, the proposed technique requires much reduced amount of training data and complexity while allowing simultaneous transmission of all active users in SC mode. This brings significant reduction in the acquisition period for slowly varying channel parameters (i.e., CCMs, JADPPs, sparsity map) by resolving the channel both in angle and delay domain. Finally, in second mode of operation, the instantaneous CSI is acquired in reduced dimensional beamspace formed by the proposed hybrid architecture exploiting the estimated power profiles.

The novelty in this paper lies behind the integration of efficient adaptive detection/estimation algorithms in radar literature with massive MIMO hybrid beamforming in order to extract the scatter map of frequency-selective multi-user channel where the MPCs are resolvable both in angular and temporal domain. The contribution is twofold. First, to the author's knowledge, there is no such prior work, that obtains the sparsity map both in angle and delay domain via CFAR thresholding for wideband SC transmission. Here, with the help of proposed methodology, which is completely different than the aforementioned studies in the literature, the sparsity map and CCMs are acquired with lowered dimensional observation after hybrid beamforming. Second, based on the estimated CCMs, a novel statistical analog beamformer, suppressing both inter-group interference and ISI, and taking the doubly sparse structure of wideband channel into account, is designed for JSDM framework by inspiring from the work in \cite{Guvensen16} (where a nearly optimal Capon like beamformer for general rank signal model was constructed). While designing the statistical pre-beamformer, the distribution of RF chains among different MPCs is optimized for SC transmission by considering the amount of interference they are subject to. After reducing the dimension via the proposed statistical beamformer, efficient beamspace aware instantaneous CSI estimators are provided. It is shown that the proposed CCM construction and reduced rank channel acquisition techniques necessitate considerably lowered slow and fast time training overhead. Furthermore, the performance benchmark for the estimation of doubly sparse SC wideband massive MIMO channel in mean square error (MSE) sense is achieved via the proposed algorithms which can be regarded as promising beam and channel acquisition techniques in this regard for next generation wireless networks. 

\textit{Notations:} Vectors and matrices are denoted by boldface small and capital letters; the transpose, Hermitian and inverse of the matrix $\mathbf{A}$ are denoted by $\mathbf{A}^T$, $\mathbf{A}^H$ and $\mathbf{A}^{-1}$; $\left[\mathbf{A}\right]_{(i,j)}$ is the $(i,j)^{th}$ entry of $\mathbf{A}$; the entry index of the vector and the matrix starts from 0; $\operatorname{Tr}\left\{\mathbf{A}\right\}$ is the trace of $\mathbf{A}$; $\det\left[\mathbf{A}\right]$ is the determinant of $\mathbf{A}$; $\mathbf{I}$ is the identity matrix with appropriate size; $\mathbb{E}\left\{\cdot\right\}$ is the statistical expectation; $|\mathcal{S}|$ denotes the cardinality of the set $\mathcal{S}$; $||\mathbf{a}||$ denotes the Euclidean norm of $\mathbf{a}$, and $\delta_{nn'}$ is the Kronecker-Delta function which is equal to $1$ if $n=n'$ otherwise $0$.

\section{System Model} 
\label{sec:sym_model}

\begin{figure}[tb]
\includegraphics[width=0.5\linewidth]{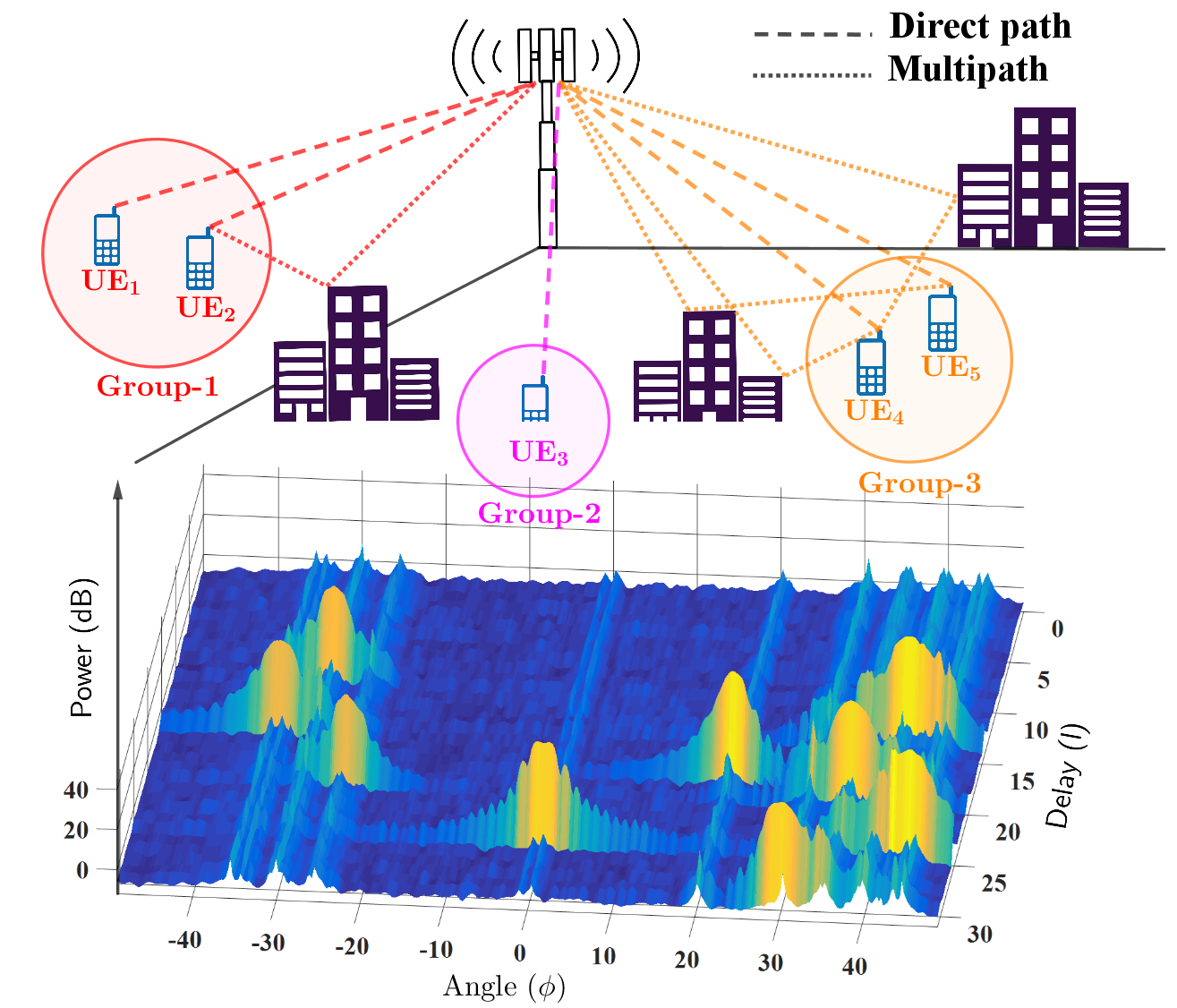}\hfill
\includegraphics[width=0.5\linewidth]{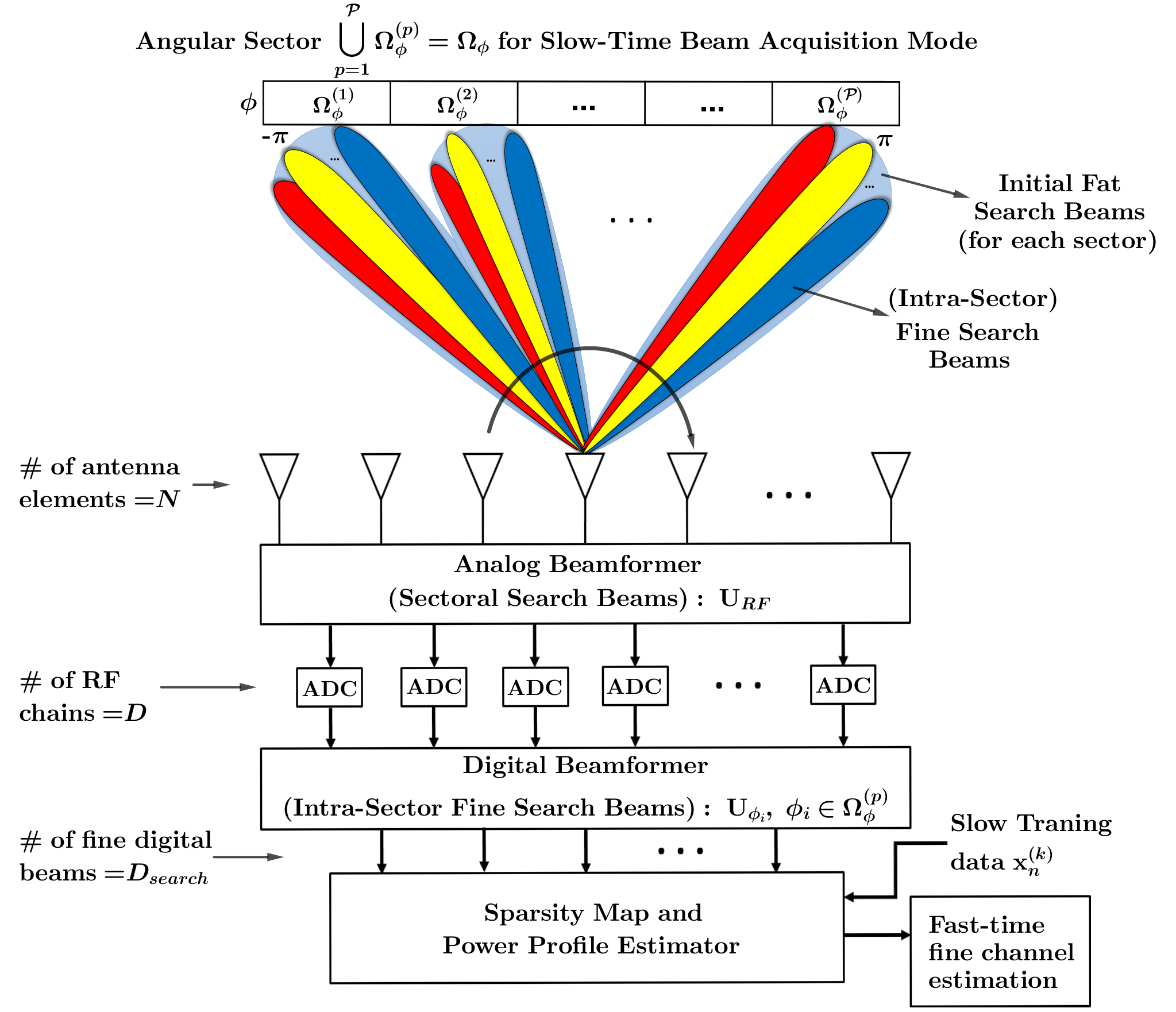}
{\phantomsubcaption\label{fig:genel_figur}}
{\phantomsubcaption\label{fig:initial_beam}}
\caption{Angle-delay power profile of active users in massive MIMO system (left) and beam acquisition scheme (right)}
\label{fig:overall_figures}
\end{figure}

We consider a multi-user massive MIMO system operating at mm-wave bands in TDD mode. The BS is equipped with $N$ antennas and serves $K$ single-antenna users. At the beginning of every coherence interval, all users transmit training sequences with length $T$. We assume a linear modulation (e.g., PSK or QAM) and a transmission over frequency-selective channel for all user equipments (UEs) with a slow evolution in time relative to the signaling interval (symbol duration). Under such conditions, the baseband equivalent received signal samples, taken at symbol rate ($W$) after pulse matched filtering, are expressed as%
\begin{equation}
\mathbf{y}_n=\sum_{k=1}^{K}\sum_{l=0}^{L-1}\mathbf{h}_l^{(k)}x_{n-l}^{(k)} + \mathbf{n}_n
\label{eqn:multiuser_multipath_mimo_model}
\end{equation}
\noindent for $n=0,\ldots,T-1$, where $\mathbf{h}_l^{(k)}$ is $N \times 1$ multipath channel vector, namely, the array impulse response of the serving BS stemming from the $l^{th}$ MPC of $k^{th}$ user. Here, $\left\{x_n^{(k)};\;-L+1 \leq n \leq T-1\right\}$ are the training symbols for the $k^{th}$ user, $L$ is the channel memory of $k^{th}$ user multipath channels. The $L-1$ symbols at the start of the preamble, prior to the first observation at $n=0$, are the precursors. Training symbols are selected from a signal constellation $S \in \mathbb{C}$ and $\mathbb{E}\left\{|x_n^{(k)}|^2\right\}=E_s$ is set to $1$ for all $k$. In (\ref{eqn:multiuser_multipath_mimo_model}), $\mathbf{n}_n$ are the additive complex white Gaussian noise (AWGN) vectors during uplink pilot segment with spatially and temporarily independent and identically distributed (\textit{i.i.d.}) as $\mathcal{CN}\left(\mathbf{0},N_0\mathbf{I}_N\right)$, and $N_0$ is the noise power. 

\subsection{Statistical Models for MPCs}
We assume Rayleigh-correlated MPCs where each user has channel $\mathbf{h}_l^{(k)} \sim \mathcal{CN}\left(\mathbf{0},\mathbf{R}_l^{(k)}\right)$. Then, their corresponding cross-covariance matrices can be expressed in the form of
\begin{equation}
\mathbb{E}\left\{\mathbf{h}_l^{(k)}\left(\mathbf{h}_{l'}^{(k')}\right)^H\right\}=\mathbf{R}_l^{(k)}\delta_{kk'}\delta_{ll'} 
\label{eqn:mimo_multipath_correlations}
\end{equation}
by using the uncorrelated local scattering model where all MPCs are assumed to be mutually independent according to the well-known wide sense stationary uncorrelated scattering (WSSUS) model \cite{adhikary14}, \cite{swindlehurst16}; the multipath channel vectors are uncorrelated with respect to $l$, and also mutually uncorrelated with that of the different users. The average received signal-to-noise ratio (\text{SNR}) for the $l^{th}$ MPC of $k^{th}$ user can be defined as $\text{SNR}_l^{(k)} \triangleq \frac{E_s}{N_0}\beta_l^{(k)}$ where $\beta_l^{(k)} \triangleq \operatorname{Tr}\{\mathbf{R}_l^{(k)}\}$\footnote{
It shows the average \text{SNR} after maximal ratio combining (MRC) when the beam is steered toward the angular location of $l^{th}$ MPC of $k^{th}$ user. Then, $\frac{1}{N}\frac{E_s}{N_0}\beta_l^{(k)}$ can be seen as the average received \text{SNR} at each antenna element before beamforming.}. Then, the total received SNR of $k^{th}$ user is $\frac{E_s}{N_0}\beta^{(k)}$ where $\beta^{(k)} \triangleq \sum_{l=0}^{{L}-1}\operatorname{Tr}\left\{\mathbf{R}_l^{(k)}\right\} = \sum_{l=0}^{{L}-1}\beta_l^{(k)}$. 

In mm-wave bands, an important phenomena is \textit{channel sparsity} observed both in angular and temporal domain. That is to say, most of the channel power is concentrated in a finite region on \textit{angle-delay} plane, corresponding to the interaction with physical clusters of scatterers in the real world \cite{adhikary14}. Thus the number of significant MPCs is reduced to a much lower value than that for a microwave system, and these dominant MPCs are seen by the BS under a very constrained angular range (AoA support). Then, CCM of a particular MPC is given by \cite{adhikary13}, \cite{Ma19}
\begin{equation}
\mathbf{R}_l^{(k)} \triangleq \int_{\mu_l^{(k)}-\frac{\Delta_l^{(k)}}{2}}^{\mu_l^{(k)}+\frac{\Delta_l^{(k)}}{2}}{\rho_l^{(k)}(\phi)\mathbf{u}(\phi)\mathbf{u}^H(\phi)d\phi}
\label{eq:ccm_def}
\end{equation}
where $\mathbf{u}(\phi) \in \mathbb{C}^{N \text{x} 1}$ with $||\mathbf{u}(\phi)||^2=1$ is the uniform linear array (ULA) manifold (steering) vector. The steering vector can be expressed as $\mathbf{u}(\phi) = \dfrac{1}{\sqrt{N}}\left[1, e^{j\pi\sin(\phi)}, \ldots, e^{j\pi(N-1)\sin(\phi)}\right]^T$ where the antenna spacing $d$ is the half of the signal carrier wavelength $\lambda$ for $-\pi\leq\phi<\pi$. In \eqref{eq:ccm_def}, $\Delta_l^{(k)}$ is the angular spread (AS) of $l^{th}$ MPC of $k^{th}$ user with mean look angle $\mu_l^{(k)}$ and $\rho_l^{(k)}(\phi)$ is the angular power density of $l^{th}$ MPC of $k^{th}$ user where $-\pi\leq\phi<\pi$. The angular power density $\rho_l^{(k)}(\phi)$ is non-zero if $\phi \in S_l^{(k)}$ where $S_l^{(k)}$ is the angular support set of $l^{th}$ MPC of $k^{th}$ user. Here, the support set is defined as $S_l^{(k)}\triangleq \left[\mu_l^{(k)}-\Delta_l^{(k)}{\big/}2,\mu_l^{(k)}+\Delta_l^{(k)}{\big/}2\right]$. Based on \eqref{eq:ccm_def}, it is simple to note that $\beta_l^{(k)}$ can be expressed in terms of $\rho_l^{(k)}(\phi)$ as $\beta_l^{(k)} = \operatorname{Tr}\left\{\mathbf{R}_l^{(k)}\right\} = \int_{-\pi}^{\pi}\rho_l^{(k)}(\phi)d\phi$.

As can be seen from \eqref{eq:ccm_def}, the CCM of a particular MPC is a function of the power intensity $\rho_l^{(k)}(\phi)$ which is non-zero only for particular values of $l$ among $\{0,1,\ldots,L-1\}$ and for a constrained angular range of $\phi$. These particular values on joint angle-delay plane for which $\rho_l^{(k)}(\phi)$ is significantly above the noise level, can be used to construct \textit{sparsity map}, a matrix composed of ones and zeros only. The non-zero entries of this matrix shows the temporal locations and angular supports of active MPCs for each user channel on joint angle-delay plane. The sparsity map together with power intensities are slowly varying in time as the AoA of each user signal evolves depending on the user mobility, variation rate of the scattering environment characteristics, etc. \cite{adhikary14}, \cite{you15}, \cite{caire16}, \cite{utschick05}. The rate of change of these long term parameters is much smaller than that of the actual small-scale fading process. This fact helps us design channel estimators in hybrid architecture after effectively reducing the signaling dimension via these slowly-varying parameters. 

\subsection{Equivalent Multi-Ray Channel Model for MPCs}
Our objective is to determine the regions in joint angle-delay power map where $\rho_l^{(k)}$ is non-zero, thus to estimate $\mathbf{R}_l^{(k)}$ based on \eqref{eq:ccm_def}. In order to realize this, the following practical model for the MPC of each user channel, namely $\mathbf{h}_l^{(k)}$ in \eqref{eqn:multiuser_multipath_mimo_model}, is adopted
\begin{equation}
\mathbf{h}_l^{(k)} \approx {\sqrt{\dfrac{\Delta_l^{(k)}}{P}}}{\sum_{p=0}^{P-1}\alpha_{l,p}^{(k)}\mathbf{u}(\phi_{l,p}^{(k)})}
\label{eq:h_approx}
\end{equation}
where the propagation from $l^{th}$ MPC of $k^{th}$ user to BS is composed of $P$ rays, and $\alpha_{l,p}^{(k)}$ represents the complex gain of the $p^{th}$ ray having AoA $\phi_{l,p}^{(k)} \triangleq \mu_l^{(k)} + \Delta_l^{(k)}\left(\frac{p}{P}-\frac{1}{2}\right),\ p=0,\ldots,P-1$ \cite{Xie17}, \cite{Xie18}. Since the WSSUS model is adopted, $\alpha_{l,p}^{(k)}$ satisfies the following equation
\begin{equation}
\mathbb{E}\left\{\alpha_{l,p}^{(k)}{\left[\alpha_{l,p'}^{(k)}\right]}^*\right\} = \rho_l^{(k)}(\phi_{l,p}^{(k)})\delta_{pp'}, \quad \phi_{l,p}^{(k)} \in S_l^{(k)}.
\end{equation}
Based on the given model in \eqref{eq:h_approx}, one can validate that the covariance of $\mathbf{h}_l^{(k)}$ asymptotically satisfies \eqref{eq:ccm_def} after Riemann integration when $P\rightarrow\infty$
\begin{equation}
\mathbf{R}_l^{(k)} = \mathbb{E}\left\{\mathbf{h}_l^{(k)}\left(\mathbf{h}_l^{(k)}\right)^H\right\} = \lim\limits_{P\to\infty}\sum_{p=0}^{P-1}\frac{\mathbb{E}\left\{\left|\alpha_{l,p}^{(k)}\right|^2\right\}\Delta_l^{(k)}}{P}\mathbf{u}(\phi_{l,p}^{(k)})\mathbf{u}^H(\phi_{l,p}^{(k)}) = \int_{S_l^{(k)}}{\rho_l^{(k)}(\phi)\mathbf{u}(\phi)\mathbf{u}^H(\phi)d\phi}.
\end{equation}
\section{Hybrid Beamforming Based Joint Angle-Delay Domain Power Profile (JADPP) Estimation} 
\label{sec:jadpp_hybrid}
In order to estimate the CCMs and sparsity map of each users, we need to estimate their JADPPs, i.e., $\rho_l^{(k)}(\phi)$ in \eqref{eq:ccm_def} together with their angular support $S_l^{(k)}$ first. Since hybrid beamforming structure is used, limited number of RF chains ($D$) is utilized ($D \ll N$). In hybrid structure, before estimating the power profile of each MPCs, initially, the sector of interest, in which the users are to be served, is divided into non-overlapping sub-angular sectors (which are scanned by initial search beams constructed in analog domain). Then, JADPPs and sparsity map are extracted for each user in the sector of interest. This mode of operation is called as \textit{slow-time beam acquisition mode} (or \textit{slow-time training mode}) as illustrated in Fig. \ref{fig:initial_beam}. In this mode, each UE transmits its \textit{slow-time} training sequence so that the BS estimates the angular locations of each user MPCs in TDD mode. We define $T\times 1$ pilot (training) sequence vector for $l^{th}$ MPC of $k^{th}$ user as follows
\begin{equation}
\mathbf{x}_l^{(k)} \triangleq \left[\begin{matrix}x_{-l}^{(k)} & \hdots & x_{T-1-l}^{(k)} \end{matrix} \right]^H,\ l=0,\ldots,L-1.
\label{eq:xlk}
\end{equation}
In addition, the angular search sector of interest, $\Omega_\phi$, which is defined as the ordered set of look angles $\phi$ (to which the beam is steered towards), is taken as $\Omega_\phi = \left\{\phi \mathrel{\Big|} \phi=\phi_i\ ;\ \phi_i = -\pi+i\frac{2\pi}{M},\ i=0,\ldots,M-1\right\}$. Then, the angular sub-sectors $\Omega_\phi^{(p)}$ in Fig. \ref{fig:initial_beam} are constructed such that $\bigcup\limits_{p=1}^{\mathcal{P}}\Omega_\phi^{(p)}=\Omega_\phi$. In slow-time training mode, where $\Omega_\phi^{(p)}$'s are scanned by initial search beams (constructed by the columns of analog beamformer matrix $\mathbf{U}_{RF}$ peculiar to each sub-sector in Fig. \ref{fig:initial_beam}) separately, each active UE repeats its training sequence $\left\{x_n^{(k)}\right\}_{n=-L+1}^{T-1}$ in \eqref{eqn:multiuser_multipath_mimo_model} at least $\mathcal{P}$ times so that the power profiles of all users in $\Omega_\phi$ are acquired by the BS.
\subsection{Discrete Time Spatio-Temporal Domain Signal Model for JADPP Estimation}
\label{sec:dt_signal_model}
If the selected look angle $\phi_i\in\Omega_\phi$ is in the support set $S_l^{(k)}$, and the AS is narrow enough (which is the case for mm-wave channels \cite{guvensen16_2}), we can approximate $\mathbf{h}_l^{(k)}$ by assuming $\phi_{l,p}^{(k)} \approx \phi_i$ in \eqref{eq:h_approx} as
\begin{equation}
\mathbf{h}_l^{(k)} \approx {\sqrt{\dfrac{\Delta_l^{(k)}}{P}}}{\sum_{p=0}^{P-1}\alpha_{l,p}^{(k)}\mathbf{u}(\phi_{l,p}^{(k)})} \approx \alpha_l^{(k)}\mathbf{u}(\phi_i),\ \phi_i\in\Omega_\phi \ \textrm{where } \alpha_l^{(k)} \triangleq \sum_{p=0}^{P-1}\sqrt{\frac{\Delta_l^{(k)}}{P}}\alpha_{l,p}^{(k)}.
\label{eq:h_approx2}
\end{equation}
Here, $\alpha_l^{(k)}$ can be regarded as the effective complex channel gain (reflection coefficient) of $l^{th}$ MPC for $k^{th}$ user at look angle $\phi_i$. Asymptotically, one can calculate the corresponding average channel power as 
\begin{equation}
\mathbb{E}\left\{\left|\alpha_l^{(k)}\right|^2\right\} = \lim\limits_{P\to\infty}\mathbb{E}\left\{\left|\sum_{p=0}^{P-1}\sqrt{\frac{\Delta_l^{(k)}}{P}}\alpha_{l,p}^{(k)}\right|^2\right\} = \lim\limits_{P\to\infty}\sum_{p=0}^{P-1}\frac{\Delta_l^{(k)}}{P}\rho_l^{(k)}(\phi_{l,p}^{(k)}) = \int_{S_l^{(k)}}\rho_l^{(k)}(\phi)d\phi = \beta_l^{(k)}
\label{eq:alpha_beta}
\end{equation}
when $P\rightarrow\infty$ in \eqref{eq:h_approx}, and it can be noted that if uniform power distribution is assumed, then $\beta_l^{(k)}=\rho_l^{(k)}\Delta_l^{(k)}$. During the slow-time training phase, if the BS intends to estimate the effective channel gain of the $l^{th}$ MPC for the $k^{th}$ user at look angle $\phi_i$, namely $\alpha_l^{(k)}$, it is useful to construct the following discrete time equivalent signal model in spatio-temporal domain by using \eqref{eqn:multiuser_multipath_mimo_model}, \eqref{eq:xlk}, and \eqref{eq:h_approx2}:
\begin{equation}
\mathbf{Y} \triangleq \left[\begin{array}{cccc} \mathbf{y}_0 & \mathbf{y}_1 & \cdots & \mathbf{y}_{T-1} \end{array}\right]_{N \times T} = \mathbf{h}_l^{(k)}\left(\mathbf{x}_l^{(k)}\right)^H + \mathbf{N}_l^{(k)} \approx \alpha_l^{(k)}(\phi_i)\mathbf{u}(\phi_i)\left(\mathbf{x}_l^{(k)}\right)^H + \mathbf{N}_l^{(k)}
\label{eq:Y}
\end{equation}
where $\mathbf{N}_l^{(k)}$, total interfering component to $l^{th}$ MPC of the $k^{th}$ user, is given as
\begin{equation}
\mathbf{N}_l^{(k)} \triangleq \sum_{l'=0,l' \neq l}^{L-1}\mathbf{h}_{l'}^{(k)}\left(\mathbf{x}_{l'}^{(k)}\right)^H + \sum_{k'=1,k' \neq k}^{K}\sum_{l'=0}^{L}\mathbf{h}_{l'}^{(k')}\left(\mathbf{x}_{l'}^{(k')}\right)^H + \left[\begin{array}{ccc}\mathbf{n}_0 & \cdots & \mathbf{n}_{T-1}\end{array}\right].
\label{eq:Nlk}
\end{equation}
Note that in \eqref{eq:Nlk}, $\mathbf{N}_l^{(k)}$ is composed of self interference signal stemming from the MPCs of $k^{th}$ user other than the $l^{th}$ MPC, inter-user interference signal and AWGN. Based on \eqref{eq:Y}, {\small$\mathbb{E}\left\{\left|\alpha_l^{(k)}\right|^2\right\}$} is to be estimated at preassumed spatio-temporal locations $\left\{\phi_i, l\right\}$ on joint angle-delay map for all active users. That is to say, average power of the MPCs, which are likely to exist, at each angular and temporal delay locations (for $l=0,\ldots,L-1$ and $\forall\phi_i \in \Omega_\phi$) is to be estimated for all active users.

\subsection{Initial Beam Acquisition Mode}
\label{sec:init_beam}

In hybrid beamforming architecture adopted, the initial analog beamformer matrix, $\mathbf{U}_{RF}$, is constructed to illuminate the intended sub-sector of interest $\Omega_\phi^{(p)}$ where $\phi_i \in \Omega_\phi^{(p)}$ as shown in Fig. \ref{fig:initial_beam}\footnote{
The initial analog beamformer $\mathbf{U}_{RF}$ used to estimate power profile in slow-time training mode needs to satisfy $\mathbf{u}(\phi_i) \approx \mathbf{U}_{RF}\mathbf{U}_{RF}^H\mathbf{u}(\phi_i)$ in order for intended sector to be covered properly.}. In order to maximize the coverage of the intended sector, the columns of $\mathbf{U}_{RF}$ (where $\mathbf{U}^H_{RF}\mathbf{U}_{RF}=\mathbf{I}_D$) can be obtained as the most dominant $D$ (number of RF chains) eigenvectors of the following matrix
\begin{equation}
\mathbf{R}^{(sector-p)} \triangleq \int_{\phi\in\Omega_\phi^{(p)}}\mathbf{u}(\phi)\mathbf{u}^H(\phi)d\phi.
\label{eq:Rsector_p}
\end{equation}
Here, $\mathbf{R}^{(sector-p)}$ can be regarded as the spatial autocorrelation matrix of user channels in sector $p$, and the analog beamforming via $\mathbf{U}_{RF}$ in this mode is nothing but the Karhunen-Loeve Transform (KLT) in angular domain \cite{van2004optimum}. Assuming that each user in sector-$p$ having a mean AoA uniformly distributed over the sector of interest, $\mathbf{R}^{(sector-p)}$ can also be considered as the initial CCM estimate of each user MPCs in sector-$p$. Similarly, intra-sector digital fine search beams for the look angle $\phi_i \in \Omega_\phi^{(p)}$ shown in Fig. \ref{fig:initial_beam}, namely $\mathbf{U}_{\phi_i}$'s are constructed after projection on the range space of $\mathbf{U}_{RF}$. After illuminating the intended sector $\Omega_\phi^{(p)}$ by $\mathbf{U}_{RF}$, the digital search beams in reduced dimension are steered towards $\phi_i$ at which {\small$\mathbb{E}\left\{\left|\alpha_l^{(k)}\right|^2\right\}$} in \eqref{eq:h_approx2} is to be estimated. In order to realize this, a particular angular region (patch) in $\Omega_\phi^{(p)}$ whose center is the look angle $\phi_i$, which we are interested in, is to be selected and illuminated by $\mathbf{U}_{\phi_i}$. Here, $\mathbf{U}_{\phi_i}$ can be contemplated as the $D \times D_{search}$ matrix of the eigenvectors corresponding to the largest $D_{search}$ eigenvalues of $\mathbf{R}_{\phi_i}$ which is defined as the reduced dimensional spatial autocorrelation matrix of the user channels of selected angular patch in $\Omega_\phi^{(p)}$ whose center is $\phi_i$, and $D_{search}$ is the search dimension (number of digital beams to be constructed). Then, $\mathbf{R}_{\phi_i}$ can be constructed for the mean look angle $\phi_i\in\Omega_\phi^{(p)}$ as 
\begin{equation}
\mathbf{R}_{\phi_i} = \mathbf{U}^H_{RF}\left(\int_{\phi_i-\sigma/2}^{\phi_i+\sigma/2}\mathbf{u}(\phi)\mathbf{u}^H(\phi)d\phi\right)\mathbf{U}_{RF}
\label{eq:Rphi}
\end{equation}
where $\sigma$ is the angular width of the selected patch in sector-$p$ which is simply called as the look spread. While constructing these initial intra-sector digital beams, $\mathbf{U}_{\phi_i}$'s can be normalized such that $\mathbf{U}_{\phi_i}^H\mathbf{U}_{\phi_i}=\mathbf{I}_{D_{search}}$ for proper operation. Then, we can define a hybrid beamformer matrix in slow-time training mode as $\mathbf{U}\triangleq\mathbf{U}_{RF}\mathbf{U}_{\phi_i}$ and express the signals of \eqref{eq:Y} in reduced dimensional digital beamspace as
\begin{equation}
\tilde{\mathbf{Y}}=\mathbf{U}^H\mathbf{Y},\quad \tilde{\mathbf{N}}_l^{(k)}=\mathbf{U}^H\mathbf{N}_l^{(k)},\quad \tilde{\mathbf{u}}(\phi_i)=\mathbf{U}^H\mathbf{u}(\phi_i).
\label{eq:YNtilde}
\end{equation}
Later, $\mathbf{U}_{RF}$ is to be updated for \textit{fast-time instantaneous channel acquisition} (as explained in Section \ref{sec:post_grouping}) by using the estimated CCMs (which are constructed via the JADPP and sparsity map of each user).

\subsection{Proposed JADPP Estimation Techniques}
\label{sec:sec_jadpp}
By using the reduced dimensional observations obtained after hybrid beamforming in \eqref{eq:YNtilde}, we propose efficient algorithms to estimate the average channel power {\small$\mathbb{E}\left\{\left|\alpha_l^{(k)}\right|^2\right\}$} in \eqref{eq:h_approx2} for each look angle $\phi_i\in\Omega_\phi^{(p)}$ and temporal delays $l=0,\ldots,L-1$. Here, we first conceive that $\alpha_l^{(k)}$ is to be estimated for each preassumed \textit{spatio-temporal resolution cell} which is defined as the pair $\left\{\phi_i, l\right\}$ on joint angle-delay map by using each slow-time training snapshot $\mathbf{Y}$ in \eqref{eq:Y}. We denote this estimate as $\hat{\alpha}_l^{(k)}(\phi_i)$. Then, the estimate of {\small$\mathbb{E}\left\{\left|\alpha_l^{(k)}\right|^2\right\}$} at look angle $\phi_i$, which is denoted by $\hat{\beta}_l^{(k)}(\phi_i)$, is constructed as $\hat{\beta}_l^{(k)}(\phi_i)={|\hat{\alpha}_l^{(k)}(\phi_i)|}^2$. Since {\small$\mathbb{E}\left\{\left|\alpha_l^{(k)}\right|^2\right\}$} is proportional with $\rho_l^{(k)}(\phi_i)$ from \eqref{eq:alpha_beta} when $|\Omega_\phi|=M$ is large enough, $\hat{\beta}_l^{(k)}(\phi_i)$ gives us the estimate of the angular power density at look angle $\phi_i$. We develop two different approaches in order to construct $\hat{\alpha}_l^{(k)}(\phi_i)$ for each spatio-temporal resolution cell $\left\{\phi_i, l\right\}$:

\subsubsection{Spatio-Temporal Adaptive Matched Filter (AMF)}
\label{sec:amf}
After reducing the dimension of spatio-temporal observation $\mathbf{Y}$ via hybrid beamformer in \eqref{eq:YNtilde}, the maximum likelihood (ML) estimate of non-random parameter $\alpha_l^{(k)}$ at look angle $\phi_i$ can be obtained as 
\begin{equation} 
\hat{\alpha}_l^{(k)}(\phi_i) = \frac{1}{\left\Vert\mathbf{x}_l^{(k)}\right\Vert^2}\frac{\tilde{\mathbf{u}}^H(\phi_i)\left[\mathbf{\Psi}_l^{(k)}\right]^{-1}\tilde{\mathbf{Y}}\mathbf{x}_l^{(k)}}{\tilde{\mathbf{u}}^H(\phi_i)\left[\mathbf{\Psi}_l^{(k)}\right]^{-1}\tilde{\mathbf{u}}(\phi_i)}\ \ \textrm{where}\ \ \mathbf{\Psi}_l^{(k)} = \tilde{\mathbf{Y}}\left[\mathbf{I}-\mathbf{x}_l^{(k)}\left(\mathbf{x}_l^{(k)}\right)^H{\big/}\left\Vert\mathbf{x}_l^{(k)}\right\Vert^2\right]\tilde{\mathbf{Y}}^H
\label{eq:amf_est}
\end{equation}
whose detailed derivation is given in Appendix I. While obtaining \eqref{eq:amf_est}, the spatial covariance matrix of interfering MPCs are also assumed to be unknown non-random parameters to be estimated together with $\alpha_l^{(k)}(\phi_i)$. We call this estimator as adaptive matched filter (AMF), inspired from the adaptive detection algorithm in \cite{robey92}, since a sample matrix inversion (SMI) type adaptive filtering is utilized in \eqref{eq:amf_est} to construct $\left[\mathbf{\Psi}_l^{(k)}\right]^{-1}$. The spatial autocorrelation matrix of interfering MPCs \big(given by $\mathbf{N}_l^{(k)}$ in \eqref{eq:Nlk}\big) is estimated by means of a simple temporal averaging of the columns of $\tilde{\mathbf{Y}}$. The most important difference of the proposed AMF from the conventional SMI based detectors in \cite{Brennan74}, \cite{robey92} and \cite{kelly86} is that the temporal averaging to form $\mathbf{\Psi}_l^{(k)}$ is obtained after projecting the observation signal $\tilde{\mathbf{Y}}$ on the desired signal nullspace \big(i.e., the null-space of $\left(\mathbf{x}_l^{(k)}\right)^H$ in \eqref{eq:xlk}\big) in order to eliminate the signal contamination due to desired MPC at $\left\{\phi_i, l\right\}$ for $k^{th}$ user.

\subsubsection{Spatio-Temporal Matched Filter (MF)}
\label{sec:mf}
As a special case, one can simplify \eqref{eq:amf_est} by taking $\mathbf{\Psi}_l^{(k)}$ as $\mathbf{I}_{D_{search}}$, which corresponds to assuming spatially white interfering signal to desired MPC at $\left\{\phi_i, l\right\}$. In this case, the following estimator, which is a spatio-temporal matched filter (MF) without any adaptive cancellation of interference in \eqref{eq:Y}, is given: 
\begin{equation}
\hat{\alpha}_l^{(k)}(\phi_i) = \frac{\tilde{\mathbf{u}}^H(\phi_i)\tilde{\mathbf{Y}}\mathbf{x}_l^{(k)}}{\left\Vert\tilde{\mathbf{u}}(\phi_i)\right\Vert^2\left\Vert\mathbf{x}_l^{(k)}\right\Vert^2}\cdot
\label{eq:mf_est}
\end{equation}
Then, the outputs of AMF/MF type estimators will be provided to the subsequent thresholding algorithm to construct the joint angle-delay sparsity map of each user in the sector of interest.

\section{User Activity Detection and Sparsity Map Construction via Constant False Alarm Rate (CFAR) Algorithm}
\label{sec:cfar}
Based on the estimated JADPPs, one can construct the sparsity map composed of the spatio-temporal resolution cells $\left\{\phi_i, l\right\}$ in joint angle-delay domain where $\rho_l^{(k)}(\phi_i)$ is determined to be non-zero for $\phi_i \in \Omega_\phi$ and $l=0,\ldots,L-1$. In order to construct the sparsity map, we apply two-stage adaptive thresholding onto $\hat{\beta}_l^{(k)}(\phi_i)={|\hat{\alpha}_l^{(k)}(\phi_i)|}^2$ \big(obtained via AMF/MF type preprocessing in \eqref{eq:amf_est} and \eqref{eq:mf_est}\big) by inspiring from the well known cell-averaging CFAR technique in radar literature \cite{richardsbook}. Thus, the regions where the power of MPCs concentrated on joint angle-delay map is determined (for each user). The following adaptive thresholding can be applied to each resolution cell $\left\{\phi_i, l\right\}$, called as \textit{cell-under-test (CUT)}, on joint angle-delay domain: $\hat{\beta}_l^{(k)}(\phi_i) \LRT{H_1}{H_0} \gamma$, $\phi_i \in \Omega_\phi$ and $l=0,\ldots,L-1$ where $\gamma$ is a CFAR threshold which is to be adaptively determined.

While constructing $\hat{\beta}_l^{(k)}(\phi_i)$, one can use multiple non-coherent snapshots, i.e., independent observations $\mathbf{Y}$ in \eqref{eq:Y} of the same sector obtained via the slow-time training data. Then, the random fluctuations on $\hat{\beta}_l^{(k)}(\phi_i)$ can be smoothed out by taking simple averaging over multiple snapshots:
\begin{equation}
\hat{\beta}_l^{(k)}(\phi_i)=\frac{1}{J}\sum_{j=1}^{J}\hat{\beta}_{l,j}^{(k)}(\phi_i),\ l=0,\ldots,L-1,\ \forall{\phi_i}\in\Omega_\phi
\label{eq:J}
\end{equation}
where $\hat{\beta}_{l,j}^{(k)}(\phi_i)=\abs{\hat{\alpha}_{l,j}^{(k)}(\phi_i)}^2$ is the estimated power profiles obtained from $j^{th}$ slow-time training snapshot for $j=1,\ldots,J$ where $J$ is the total number of slow-time training snapshots.

\subsection{Two-Stage CFAR Algorithm}
\label{sec:twostageCFAR}
We propose the following \textit{Two-Stage CFAR} algorithm which can be realized with adaptive thresholding both in temporal and spatial domain for the CUT on joint angle-delay domain:

\subsubsection{Temporal thresholding for selected resolution cell}
For each user, we first apply temporal thresholding on $\hat{\beta}_l^{(k)}(\phi_i)$ at each look angle $\phi_i \in \Omega_\phi$:
\begin{equation}
  \hat{\beta}_l^{(k)}(\phi_i) \LRT{H_1}{H_0}
  \gamma_1=\left(\bar{P}_{FA}^{-\frac{1}{L-1}}-1\right)\left(\sum_{l'=0, l' \neq l}^{L-1}\hat{\beta}_{l'}^{(k)}(\phi_i)\right),\ \phi_i\in\Omega_\phi
	\label{eq:test_in_temp}
\end{equation}
where $\bar{P}_{FA}$ is the desired average false alarm probability of the test. In \eqref{eq:test_in_temp}, the adaptive threshold $\gamma_1$ is obtained by simple averaging over JADPPs of each user for different delay locations other than $l$ at the selected look angle $\phi_i$ similar to the well-known cell-averaging CFAR tests in radar literature \cite{richardsbook}. By assuming that $\hat{\beta}_l^{(k)}$'s $\forall l$ are exponential \textit{i.i.d.} random variables, one can obtain the threshold level $\gamma_1$ which provides constant false alarm rate despite varying interference power levels\footnote{
The false alarm rate is nothing but the probability of declaring an empty resolution cell, where $\rho_l^{(k)}(\phi_i)=0$, as an active MPC having non-zero power level.}. That is to say, the test in \eqref{eq:test_in_temp} yields average false alarm probability which does not depend on the actual value of interfering signal levels.

\subsubsection{Spatial thresholding for selected resolution cell}
Similarly, we can apply adaptive spatial thresholding on estimated JADPPs at each spatio-temporal location $\left\{\phi_i, l\right\}$ on angle-delay domain as
\begin{equation}
  \hat{\beta}_l^{(k)}(\phi_i) \LRT{H_1}{H_0}
  \gamma_2=\left(\bar{P}_{FA}^{-\frac{1}{M-\left|\Pi_{\phi_i}\right|}}-1\right)\left({\sum_{\left\{\forall \phi' \in \Omega_\phi \vert \phi' \notin \Pi_{\phi_i}\right\}}} {\hat{\beta}_{l}^{(k)}(\phi')}\right),\ \phi_i\in\Omega_\phi
	\label{eq:test_in_space}
\end{equation}
where $M=|\Omega_\phi|$, and $\Pi_{\phi_i}$ is the guard interval, which is an angular window with mean angle of $\phi_i$. It is taken as $\Pi_{\phi_i} = \left\{\phi \mathrel{\Big|} \phi=\phi_i+m\frac{2\pi}{M},\ m=-\kappa,\ldots,\kappa\right\}$ where $\left|\Pi_{\phi_i}\right|\triangleq2\kappa+1$ is the length of guard interval, i.e., the number of angular resolution cells around the CUT (which are taken as guard cells, not used in cell averaging). In \eqref{eq:test_in_space}, the adaptive threshold $\gamma_2$ is obtained by averaging over estimated JADPPs of each user for different look angles ${\phi}^{'} \in \Omega_{\phi} \setminus \Pi_{\phi_i}$ at selected temporal delay $l$. 

\subsection{Joint Angle-Delay Domain Sparsity Map Construction of User Power Profiles}
\begin{figure}[tb]
\includegraphics[width=0.5\linewidth]{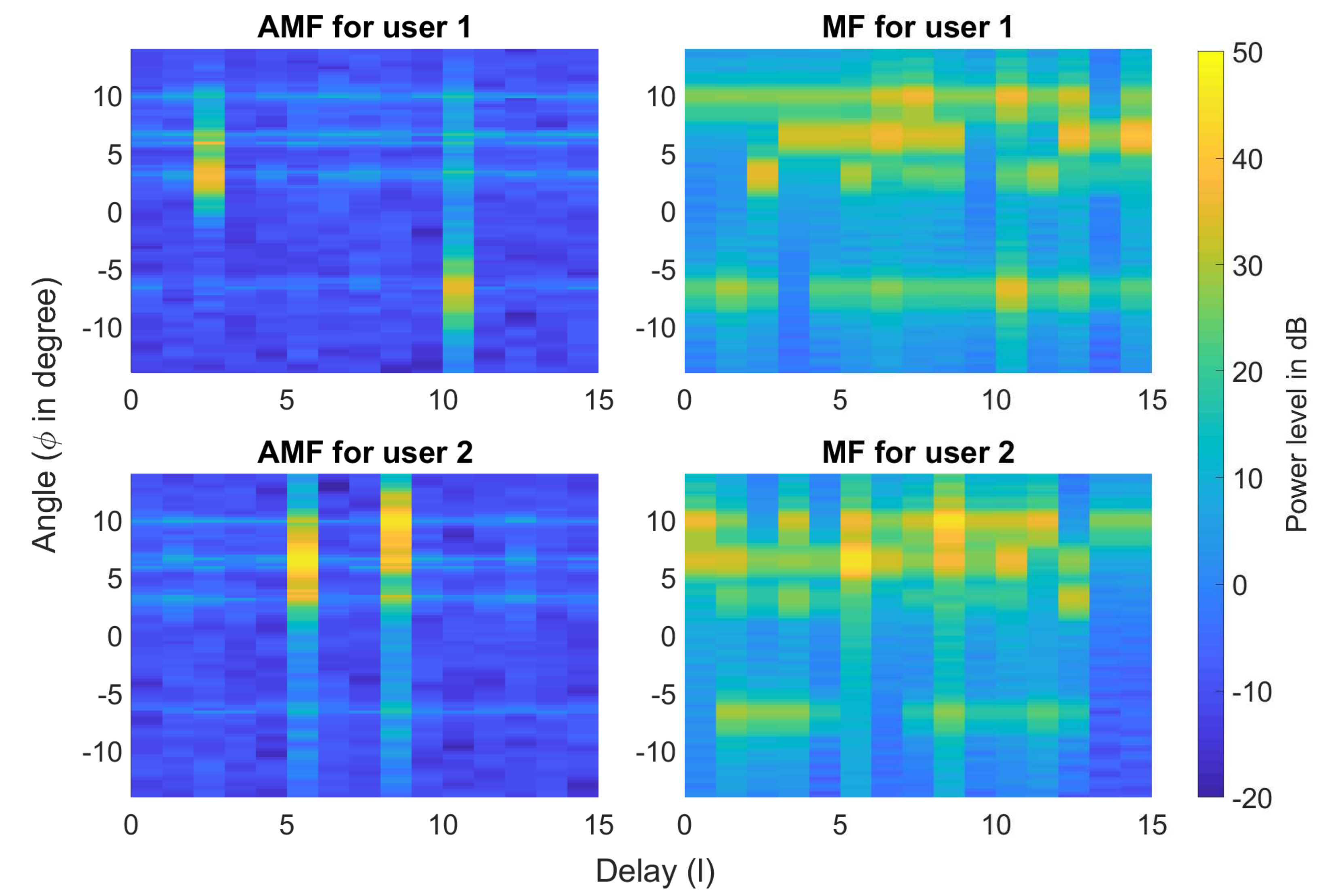}\hfill
\includegraphics[width=0.5\linewidth]{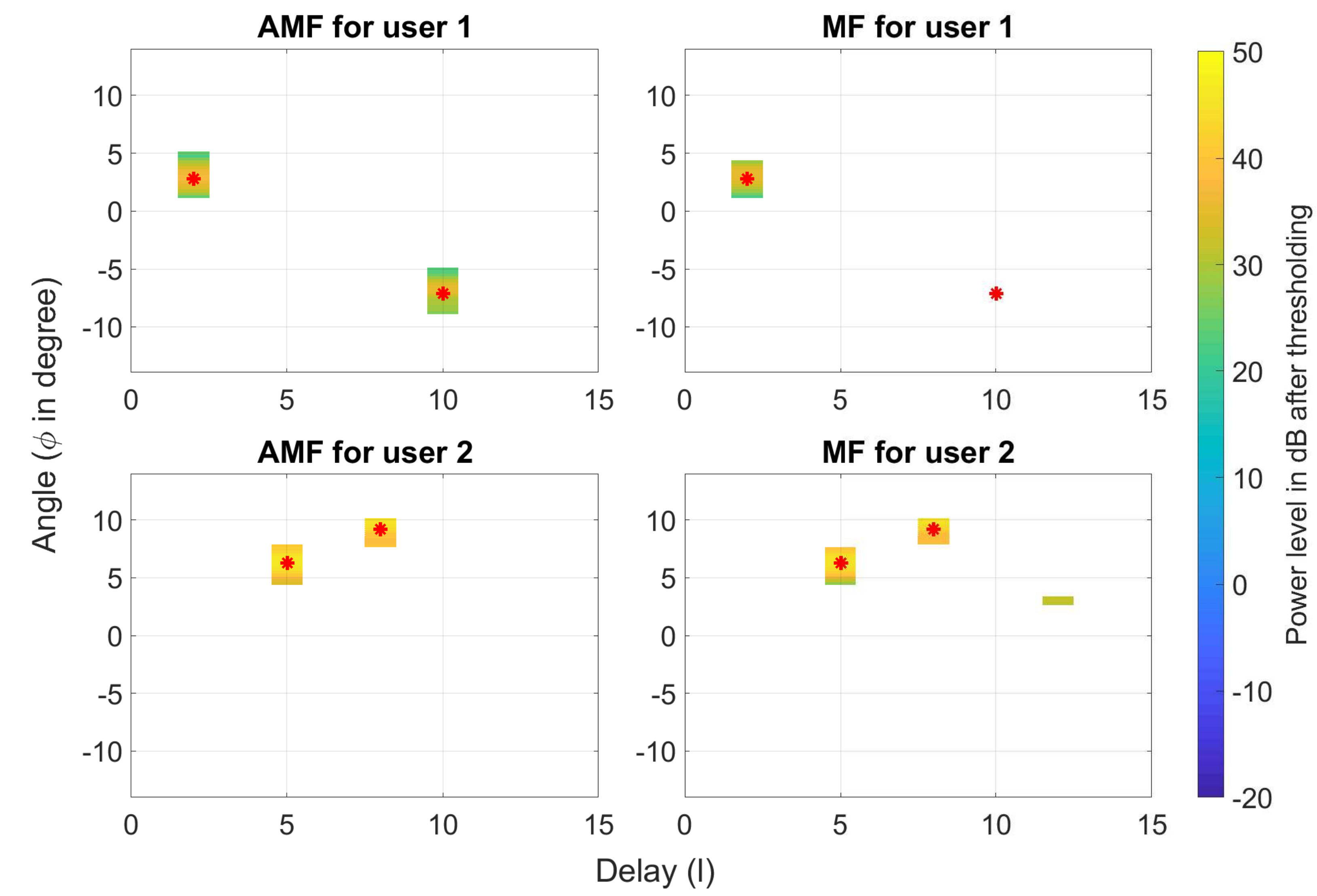}
{\phantomsubcaption\label{fig:fig_powerprofile}}
{\phantomsubcaption\label{fig:fig_sparsity}}
\caption{JADPP (left) and sparsity map (right) for user 1 and user 2 ($K=2$, $T=16$, $\beta^{(1)}=40$ dB, $\beta^{(2)}=50$ dB)}
\label{fig:jadpp_sparsity}
\end{figure}

After thresholding, the sparsity map of $k^{th}$ user, $\mathbf{I}^{(k)}$, which is $M \times L$ matrix, where $M$ is the number of resolution cells in angular domain, is constructed. If there is a detection at a particular CUT, the corresponding entry is set to 1, otherwise 0. That is to say for $l=0,\ldots,L-1$ and $\forall\phi_i\in\Omega_\phi$:
\begin{equation}
  \left[\mathbf{I}^{(k)}\right]_{(i,l)}=
  \begin{cases}
    1, & \text{if}\ \ \hat{\beta}_l^{(k)}(\phi_i)>\max{\{\gamma_1,\gamma_2\}}\\
    0, & \text{otherwise}
  \end{cases}.
	\label{eq:sparsity_map}
\end{equation}
Note that, when all elements of $\mathbf{I}^{(k)}$ is zero, it means that $k^{th}$ user is not an active user.\\
\noindent \underline{\textit{An Exemplary Scenario:}}
In Fig. \ref{fig:fig_powerprofile}, we demonstrate the estimated power levels of each MPCs of active users via the proposed JADPP estimators, namely AMF and MF. We investigate a scenario where a BS with $N=100$ antenna elements in the form of ULA is serving $2$ single-antenna users. It is assumed that both users have two active MPCs where the resolution cells with non-zero power levels are $\left\{\phi, l\right\}=\{2.8, 2\}$ and $\{-7.1, 10\}$ for the first user, and $\left\{\phi, l\right\}=\{6.3, 5\}$ and $\{9.2, 8\}$ for the second one respectively. For all MPCs, the AS is taken as $2\degree$ and $T=16$ is used in simulations. It is assumed that the received power levels are $\beta^{(1)}=40$ dB and $\beta^{(2)}=50$ dB. In Fig. \ref{fig:fig_sparsity}, we provide sparsity maps after applying two-stage thresholding on JADPP estimations given in Fig. \ref{fig:fig_powerprofile}. Here, $\bar{P}_{FA}$ is set to $10^{-3}$ and $\left|\Pi_{\phi_i}\right|$ is taken as $4\degree$.

\subsection{Performance Metrics for User Activity Detection}
Here, our aim is to find the probability of detecting the active MPCs, which have non-zero power at a given angle-delay resolution cell. We use two performance metrics to compare AMF in \eqref{eq:amf_est} and MF in \eqref{eq:mf_est}, namely the probability of detection ($P_D$) and the probability of false alarm ($P_{FA}$). By using the constructed sparsity map in \eqref{eq:sparsity_map}, \textit{the probability of detection} for the active MPCs of $k^{th}$ user can be expressed as
\begin{equation}
P_D^{(k)} \triangleq \frac{1}{\lvert\mathcal{L}^{(k)}\rvert}\sum_{l \in \mathcal{L}^{(k)}}\Pr\left\{\left[\mathbf{I}^{(k)}\right]_{(\psi_l^{(k)},l)}=1\right\}
\label{eq:pd_def}
\end{equation}
\noindent where $\mathcal{L}^{(k)}$ is the set of indices of active MPCs having positive $\beta_l^{(k)}$, and $\psi_l^{(k)}$ is the angular index pointing the mean AoA for $l^{th}$ MPC of $k^{th}$ user: $\psi_l^{(k)} \triangleq \underset{i=0,\ldots,M-1}{\operatorname*{arg\,min}} \lvert{\mu_l^{(k)}-\Omega_\phi(i)}\rvert$ where $\Omega_\phi(i)$ is $i^{th}$ element of $\Omega_\phi$. Similarly, \textit{the probability of false alarm}, showing the probability that inactive MPCs (having zero $\beta_l^{(k)}$) are declared as active for $k^{th}$ user, can be expressed as
\begin{equation}
P_{FA}^{(k)} \triangleq \frac{1}{ML}\sum_{l=0}^{L-1}\sum_{i=0,i \notin \Gamma_l^{(k)}}^{M-1}\Pr\left\{\left[\mathbf{I}^{(k)}\right]_{(i,l)}=1\right\}
\end{equation}
where $\Gamma_l^{(k)}$ is the set of indices $i$ such that the look angle $\phi_i\in\Omega_\phi$ is inside the angular support set $S_l^{(k)}$. It can be given as $\Gamma_l^{(k)}=\left\{i\mathrel{\Big|}i=0,\ldots,M-1 \textrm{ and } \Omega_\phi(i)\in S_l^{(k)}\right\}$.

\section{Sparse CCM Construction} 
\label{sec:ccm_construct}
We can construct CCMs purely based on the estimated power levels and sparsity map for each user. It is important to note that the observation signal $\mathbf{Y}$ in \eqref{eq:Y} is not available in full dimension in our hybrid structure, however each CCM needs to be estimated in full dimension. Also, we need to obtain accurate enough CCM estimates with significantly reduced amount of training snapshots. That is why we need parametric construction of CCM by exploiting its reduced rank property due to sparse nature of the channel. Hence, we can construct CCMs for each user by using sparsity matrix $\mathbf{I}^{(k)}$ and power estimates $\hat{\beta}_l^{(k)}$ as follows
\begin{equation}
  \hat{\mathbf{R}}_l^{(k)}=
  \begin{cases}
    \mathbf{0}_{N\text{x}N}, & \text{if}\ c_l^{(k)}=0 \\
    \displaystyle{\sum_{i=0}^{M-1}}\frac{\hat{\beta}_l^{(k)}(\phi_i)}{c_l^{(k)}}[\mathbf{I}^{(k)}]_{(i,l)}\mathbf{u}(\phi_i)\mathbf{u}^H(\phi_i), & \text{otherwise}
  \end{cases}
	\qquad\textrm{where } c_l^{(k)} \triangleq \sum_{i=0}^{M-1}[\mathbf{I}^{(k)}]_{(i,l)}
	\label{eq:est_ccm_def}
\end{equation}
\noindent for $l=0,\ldots,L-1$ and $\phi_i\in\Omega_\phi$. The proposed construction technique in \eqref{eq:est_ccm_def} is completely different from the conventional ones, which are based on SMI type temporal averaging in full dimension necessitating large amount of training snapshots for proper operation \cite{Brennan74}.

\noindent \underline{\textit{Asymptotic Convergence of the Proposed CCM Estimation:}}\\ 
It is interesting to see that $\hat{\mathbf{R}}_l^{(k)}$ in \eqref{eq:est_ccm_def} converges to true CCM values, $\mathbf{R}_l^{(k)}$ in \eqref{eq:ccm_def} when the sparsity map in \eqref{eq:sparsity_map} is perfectly acquired, i.e., $\left[\mathbf{I}^{(k)}\right]_{(i,l)}=1$ if $\phi_i\in S_l^{(k)}$ and $\left[\mathbf{I}^{(k)}\right]_{(i,l)}=0$ otherwise. When $\hat{\beta}_l^{(k)}(\phi_i)=\rho_l^{(k)}(\phi_i)\Delta_l^{(k)}$ (if AS is narrow enough), we can obtain the asymptotic value of $\hat{\mathbf{R}}_l^{(k)}$ by letting $c_l^{(k)}=P$ in \eqref{eq:est_ccm_def}, and $M,P\rightarrow\infty$ as follows
\begin{align}
\hat{\mathbf{R}}_l^{(k)} &= \lim\limits_{M\to\infty}\sum_{\left\{\substack{\forall\phi_i \in S_l^{(k)}\\i=0,\ldots,M-1}\right\}}\frac{\rho_l^{(k)}(\phi_i)}{P}\Delta_l^{(k)}\mathbf{u}(\phi_i)\mathbf{u}^H(\phi_i) \nonumber \\
&= \lim\limits_{P\to\infty}\frac{\Delta_l^{(k)}}{P}\sum_{p=0,\phi_{l,p}^{(k)}\in S_l^{(k)}}^{P-1}\rho_l^{(k)}(\phi_{l,p}^{(k)})\mathbf{u}(\phi_{l,p}^{(k)})\mathbf{u}^H(\phi_{l,p}^{(k)}) \quad \left(\frac{\Delta_l^{(k)}}{P} \rightarrow d\phi,\ \phi_{l,p}^{(k)} \rightarrow \phi \right) \nonumber \\
&= \int_{S_l^{(k)}}{\rho_l^{(k)}(\phi)\mathbf{u}(\phi)\mathbf{u}^H(\phi)d\phi} = \mathbf{R}_l^{(k)}.
\end{align}
\noindent This shows that when the number of angular resolution cells is high enough, and accurate JADPP estimates are available, parametric construction of $\mathbf{R}_l^{(k)}$ by \eqref{eq:est_ccm_def} in full dimension can be realized efficiently after hybrid beamforming.

\section{Nearly Optimal Covariance-Based Reduced Rank Hybrid Beamformer Design}
\label{sec:beamformer}
In this section, based on the estimated CCMs in \eqref{eq:est_ccm_def}, the analog beamformer, $\mathbf{U}_{RF}$ in Fig. \ref{fig:initial_beam} can be updated by using the statistical properties of each user channels. In \textit{slow-time beam acquisition} mode, the long term parameters of each user channels, i.e., JADPPs $\rho_l^{(k)}(\phi)$, the sparsity map $\mathbf{I}^{(k)}$, and CCMs $\mathbf{R}_l^{(k)}$ are acquired by using slow-time training snapshots as explained in previous chapters. Also, the analog beamformer $\mathbf{U}_{RF}$ can be optimized by exploiting these slowly-varying long term parameters. It is important to note that slow-time training data is transmitted much slower compared to the rate of change of instantaneous CSI.

\subsection{User Grouping Stage}
In slow-time beam acquisition mode, since no apriori information is assumed related with these slowly varying parameters initially, predetermined search sector beams together with slow-time training data are utilized to extract the spatial signatures of each user \big(based on AMF in \eqref{eq:amf_est} and MF in \eqref{eq:mf_est}\big). This initial step can be regarded as \textit{pre-grouping stage}. After this initial stage, an efficient reconstruction of analog beamformer $\mathbf{U}_{RF}$ in Fig. \ref{fig:initial_beam} can be carried out by using the estimated CCMs and sparsity map. This can be realized by means of a virtual sectorization via second-order channel statistics based \textit{user-grouping} inspired from JSDM framework \cite{adhikary13}, \cite{nam14}. In this framework, all active $K$ users can be divided into $G$ groups based on their spatial information, i.e., the estimated sparsity map and CCMs, by using proper user grouping algorithms known in the literature\footnote{
The design of user grouping algorithm is out of scope of this paper. An efficient procedure can be found in \cite{nam14}, \cite{Gesbert16}.} (as in the case of JSDM). We define $\Omega_g$ as the set of all UEs belonging to group $g$ with cardinality $|\Omega_g|=K_g$, and $\left\{g_k\right\}_{k=1}^{K_g}$ are UE indices forming $\Omega_g$, where the $K_g$ users in group $g$ are assumed to have statistically \textit{i.i.d.} channels.\\
\indent In user grouping stage, one need to construct the common covariance matrix of each MPCs in group $g$, denoted by $\mathbf{R}_l^{(g)}$ which can be considered as the common spatial covariance matrix of UEs belonging to group $g$ at $l^{th}$ delay. Instead of using the true covariance matrices of each MPCs, which can not be known accurately, we can construct estimated $\mathbf{R}_l^{(g)}$ by using the acquired CCMs in \eqref{eq:est_ccm_def} as follows
\begin{equation}
\hat{\mathbf{R}}_l^{(g)} \triangleq \sum_{k=1}^{K_g}\hat{\mathbf{R}}_l^{(g_k)},\ g=1,\ldots,G
\label{eqn:group_delay_ccm}
\end{equation}
where $\hat{\mathbf{R}}_l^{(g_k)}$ is the estimated covariance matrix for the $l^{th}$ MPC of the $k^{th}$ user in group $g$. Similarly, the estimated spatial covariance matrix of received signal in \eqref{eqn:multiuser_multipath_mimo_model} (assuming that the transmitted symbols are \textit{i.i.d.} with unity power) can be obtained as
\begin{equation}
\hat{\mathbf{R}}_y \triangleq \displaystyle\sum_{g=1}^{G}\sum_{l=0}^{L-1}\hat{\mathbf{R}}_l^{(g)}+N_0\mathbf{I},
\label{eqn:group_received_ccm}
\end{equation}
and the estimated covariance matrix of the inter-group interference to group $g$, consisting of the statistical information for all inter-group users interfering with group $g$, can be constructed as
\begin{equation}
\hat{\mathbf{R}}_{\eta}^{(g)} \triangleq \hat{\mathbf{R}}_y-\sum_{l=0}^{L-1}\hat{\mathbf{R}}_l^{(g)},\ g=1,\ldots,G.
\label{eqn:group_interference_ccm}
\end{equation}

\subsection{Post-User Grouping Stage}
\label{sec:post_grouping}
After user grouping stage, an efficient analog beamformer for each user group can be designed via the estimated group covariance matrices given in \eqref{eqn:group_delay_ccm}, \eqref{eqn:group_received_ccm}, \eqref{eqn:group_interference_ccm}. This stage can be regarded as \textit{post-grouping stage}. Here, we load $\mathbf{U}_{RF}$ in Fig. \ref{fig:initial_beam} with the optimized statistical analog beamformer, which is applied in order to distinguish \textit{intra-group} signal of users in group $g$ from other groups by suppressing the \textit{inter-group interference} while reducing the signaling dimension of $\mathbf{Y}$ in \eqref{eq:Y}. In this stage, a $D_gT$-dimensional space-time vector $\mathbf{y}^{(g)}$, where $D_g$ is the number of RF chains assigned to group $g$, can be formed by using $\mathbf{Y}$ for all groups after the following linear transformation:
\begin{equation}
\mathbf{y}^{(g)} \triangleq 
\left(\mathbf{I}_T \otimes \left[\mathbf{S}^{(g)}\right]^H\right)\operatorname{vec}\left\{\mathbf{Y}\right\},\ g=1,\ldots,G
\label{eq:space_time_noise_reduced}
\end{equation}
where $\mathbf{S}^{(g)}$ is an $N \times D_g$ statistical analog beamformer matrix that projects the $N$-dimensional received signal samples $\left\{\mathbf{y}_n\right\}_{n=0}^{T-1}$ in (\ref{eqn:multiuser_multipath_mimo_model}) on a suitable $D_g$-dimensional subspace in spatial domain. Here, since a limited number of RF chains, $D$, are used in hybrid architecture, we have the following constraint: $\sum_{g=1}^{G}D_g=D$. After constructing the optimal $\mathbf{S}^{(g)}$, $\mathbf{U}_{RF}$ in Fig. \ref{fig:initial_beam} is replaced with $\left[\begin{array}{cccc} \mathbf{S}^{(1)} & \mathbf{S}^{(2)} & \cdots & \mathbf{S}^{(G)} \end{array}\right]_{N\times D}$.

\subsubsection{MPC Grouping for Efficient Analog Beamformer Design}
If there exists a significant overlap among some of the MPCs of group $g$ in the angular domain, one can simply form groups of nonresolvable MPCs, and process them jointly. One can group $l^{th}$ and ${l^{'}}^{th}$ active MPCs of group $g$ into one MPC group if the following criteria holds
\begin{equation}
\varphi_{ll^{'}}^{(g)} \triangleq \frac{|\Gamma_l^{(g)} \cap \Gamma_{l^{'}}^{(g)}|}{\operatorname{min}\left\{|\Gamma_l^{(g)}|,|\Gamma_{l^{'}}^{(g)}|\right\}} \geq \zeta
\label{eqn:angular_resolvability}
\end{equation}
where $\zeta$ is the overlapping threshold, and $\Gamma_l^{(g)}$, $\Gamma_{l^{'}}^{(g)}$ are the set of indices $i$ such that $\phi_i\in\Omega_\phi$ is inside the angular region which is determined to have non-zero power level for $l^{th}$ and ${l^{'}}^{th}$ MPCs of group $g$ respectively:
\begin{equation}
\Gamma_l^{(g)}\triangleq \left\{i\mathrel{\Big|}\sum_{k=1}^{K_g}\left[\mathbf{I}^{(g_k)}\right]_{(i,l)}\neq0,\ g_k\in \Omega_g, i=0,\ldots,M-1\right\}.
\end{equation}
If (\ref{eqn:angular_resolvability}) is not satisfied, the corresponding MPCs are assumed to be \textit{resolvable} in angular domain. After computing $\varphi_{ll^{'}}^{(g)}$ for all $(l,l^{'})^{th}$ active pair, one can group nonresolvable active MPCs to form spatially resolvable MPC groups in angular domain which is denoted by $\mathfrak{L}_{\ell}^{(g)}$:
\begin{equation} 
\mathfrak{L}_{\ell}^{(g)} = \{l\ |\ \textrm{a set of temporal indices of spatially overlapping active MPCs in user group } g\}.
\label{eq:resolvable_mpc_groups}
\end{equation}
Here, $\ell$ shows the indices of MPC groups consisting of non-resolvable components\footnote{
One can denote the set of temporal indices of all (resolvable and nonresolvable) active MPCs of group $g$ as $\mathcal{L}^{(g)}$, and the total number of resolvable MPCs of group $g$ in angular domain as $\mathcal{MPC}^{(g)}$ after MPC grouping. Then, it can be seen that $\mathcal{MPC}^{(g)} \leq |\mathcal{L}^{(g)}| \leq L$, and $\sum_{\ell=0}^{\mathcal{MPC}^{(g)}-1}|\mathfrak{L}_{\ell}^{(g)}|=|\mathcal{L}^{(g)}|$.}. Note that it is enough for an active MPC to be put into an MPC group if it has significant overlap with at least one of the MPCs in that group. In other words, $l^{th}$, ${l^{'}}^{th}$, and ${l^{''}}^{th}$ MPCs are put into the same MPC group even if $\varphi_{ll^{''}}^{(g)} < \zeta$ but $\varphi_{ll^{'}}^{(g)} \geq \zeta$, $\varphi_{l^{'}l^{''}}^{(g)} \geq \zeta$. Then, the CCM of $\mathfrak{L}_{\ell}^{(g)}$ can be constructed as $\hat{\mathbf{R}}_{\ell}^{(g)} \triangleq \sum_{l\in\mathfrak{L}_{\ell}^{(g)}}\hat{\mathbf{R}}_l^{(g)}$.

\subsubsection{Nearly Optimal Analog Beamformer Construction}
\label{sec:abeam_design}
By assuming that eigenspaces of each MPCs are nearly orthogonal, which is indeed the case in mm-wave massive MIMO systems due to sparse nature of the channel, we can construct the analog beamformer of group $g$ as $\mathbf{S}^{(g)} \triangleq \left[\begin{array}{cccc} \mathbf{S}_0^{(g)} & \mathbf{S}_1^{(g)} & \cdots & \mathbf{S}_{\mathcal{MPC}^{(g)}-1}^{(g)} \end{array}\right]_{N \times D_g}$ where the $N \times d_\ell^{(g)}$ sub-matrix $\mathbf{S}_{\ell}^{(g)}$ can be seen as the sub-beamformer that allows $\ell^{th}$ resolvable MPC of group $g$ to pass while suppressing the \textit{inter-group interference} in the spatial domain, and also the rejecting each MPC of group $g$ other than the one at delay $\ell$. Here, $d_\ell^{(g)}$ is the number of RF chains assigned to the $\ell^{th}$ MPC of group $g$ where $\sum_{\ell=0}^{\mathcal{MPC}^{(g)}-1}d_\ell^{(g)}=D_g$. For a given $d_\ell^{(g)}$, nearly optimal analog beamformer matrix, $\mathbf{S}_{\ell}^{(g)}$, can be obtained as
\begin{equation}
\mathbf{S}_{\ell}^{(g)} \triangleq \eigs(\hat{\mathbf{R}}_\ell^{(g)},\hat{\mathbf{R}}_y-\hat{\mathbf{R}}_\ell^{(g)},d_\ell^{(g)})=\eigs(\hat{\mathbf{R}}_\ell^{(g)},\hat{\mathbf{R}}_y,d_\ell^{(g)})
\label{eq:S_ell_g}
\end{equation}
where $\eigs$ operation yields $d_\ell^{(g)}$ dominant generalized eigenvectors of $\hat{\mathbf{R}}_\ell^{(g)}$ and $\hat{\mathbf{R}}_y$ corresponding to largest generalized eigenvalues. The procedure, here, is adopted from \cite{Guvensen16} by replacing the true group CCMs with the estimated ones.

\subsubsection{Optimal RF Chain Distribution Among MPCs}
First, we assume that RF chains are distributed to groups proportional with the number of users they have. Then, the optimal values of $\{d_\ell^{(g)}, \ell=0,\ldots,\mathcal{MPC}^{(g)}\scalebox{0.8}[1.0]{\( - \)}1\}$, in terms of channel estimation accuracy, can be found by using the generalized eigenvalues as
\begin{equation}
\{d_\ell^{(g)}\}_{opt} = \operatorname*{arg\,min}_{\{d_\ell^{(g)}\}} \sum_{\ell=0}^{\mathcal{MPC}^{(g)}-1}\sum_{n=1}^{d_\ell^{(g)}} \frac{1}{\lambda_{\ell,n}^{(g)}+1}
\label{eqn:rf_chain}
\end{equation}
\noindent with the constraint of $\sum_{\ell=0}^{\mathcal{MPC}^{(g)}-1}d_\ell^{(g)}=D_g$, and $\sum_{g=1}^{G}D_g=D$ \cite{Guvensen16}. Here, $\lambda_{\ell,n}^{(g)}$ is the $n^{th}$ dominant generalized eigenvalue of $\hat{\mathbf{R}}_\ell^{(g)}$ and $\hat{\mathbf{R}}_{\eta^{'}}^{(g)}\triangleq\hat{\mathbf{R}}_y-\hat{\mathbf{R}}_\ell^{(g)}$ for $\ell=1,\ldots,\mathcal{MPC}^{(g)}-1$. It can be shown from \eqref{eq:S_ell_g} that
\begin{equation}
\lambda_{\ell,n}^{(g)} = \left[\left(\left[\mathbf{S}_\ell^{(g)}\right]^H\hat{\mathbf{R}}^{(g)}_{\eta^{'}}\mathbf{S}_\ell^{(g)}\right)^{-1}\left(\left[\mathbf{S}_\ell^{(g)}\right]^H \hat{\mathbf{R}}_\ell^{(g)}\mathbf{S}_\ell^{(g)}\right)\right]_{(n,n)},\ n=1,\ldots,d_\ell^{(g)}.
\end{equation}
Various other criteria different than (\ref{eqn:rf_chain}) can also be used to distribute RF chains among different active MPCs as stated in \cite{Guvensen16}, which head to same optimal distribution.

\section{Instantaneous Channel Estimation via Hybrid Beamforming}
\label{sec:chan_est}
After slow-time beam acquisition mode and user grouping based on JSDM framework, fine instantaneous channel estimates in reduced dimension can be obtained for each user. We name this acquisition stage as \textit{fast-time acquisition mode} where reduced rank instantaneous CSI estimation can be carried out accurately and efficiently in a proper beamspace (formed by $\mathbf{S}^{(g)}$) via small amount of fast-time training data in front of each data transmission. In this mode, with the help of updated analog beamformer, the training overhead is shown to be reduced considerably.

The subsequent downlink or uplink processing, preceded by the analog beamforming, can access and utilize only the \textit{effective} multipath channel vector of each group user. The effective channels of each user, appearing at the output of analog beamformer, (which is utilized by digital beamformer for intra-group processing) can be defined as $\mathbf{h}_{eff,l}^{(g_k)} \triangleq \left[\mathbf{S}^{(g)}\right]^H\mathbf{h}_l^{(g_k)}$ where $\mathbf{h}_l^{(g_k)}$ is the full dimensional channel vector for the $l^{th}$ MPC of the $k^{th}$ user in group $g$. Then, it will be beneficial to express the variables in a single concatenated vector, namely, the \textit{effective} extended multipath channel vector of group $g$, as
\begin{equation}
\bar{\mathbf{h}}_{eff}^{(g)}\!\triangleq\!\left[\left[\bar{\mathbf{h}}_{eff}^{(g_1)}\right]^{H}\,\left[\bar{\mathbf{h}}_{eff}^{(g_2)}\right]^{H}\!\hdots\,\left[\bar{\mathbf{h}}_{eff}^{(g_{K_g})}\right]^{H}\right]^{H}\!\!,\ \bar{\mathbf{h}}_{eff}^{(g_k)}\!\triangleq\!\left[\left[\mathbf{h}_{eff,0}^{(g_k)}\right]^{H}\,\left[\mathbf{h}_{eff,1}^{(g_k)}\right]^{H}\!\hdots\,\left[\mathbf{h}_{eff,L-1}^{(g_k)}\right]^{H}\right]^{H}\!\!.
\label{eq:h_eff_bar}
\end{equation}
\indent In this mode, a $T_{fast} \times L$ training matrix (or convolution matrix), comprising of the transmitted pilots with the precursors for $k^{th}$ user in group $g$, is defined as $\mathbf{X}_k^{(g)}\triangleq\conj\left\{\left[\begin{matrix} \mathbf{x}_0^{(g_k)} & \mathbf{x}_1^{(g_k)} & \hdots & \mathbf{x}_{L-1}^{(g_k)} \end{matrix} \right]\right\}$, where $\mathbf{x}_l^{(g_k)}$ is defined in \eqref{eq:xlk}. The length of fast-time training data is taken as $T_{fast}$ which is supposed to be much smaller than that of the slow-time training data $T$. Then, the complete fast-time training matrix of all users in group $g$ during the signaling interval $T_{fast}$ is given by\footnote{
Here, asynchronous transmission is possible. We consider a fast-time training phase in which only the intended group $g$ trains and other groups are in the data mode where their transmitted symbols are assumed to be \textit{i.i.d.}. Alternatively, same fast-time training matrix can be assigned to different groups since different groups are discriminated by their spatial signatures via the proposed analog beamformer.} 
\begin{equation}
\mathbf{X}^{(g)}\triangleq \left[\begin{array}{cccc}
\mathbf{X}_1^{(g)} & \mathbf{X}_2^{(g)} & \cdots & \mathbf{X}_{K_g}^{(g)}
\end{array}\right]_{T_{fast} \times K_gL}. 
\label{eqn:complete_training_matrix}
\end{equation}
By using the definitions in \eqref{eq:h_eff_bar} and \eqref{eqn:complete_training_matrix} together with \eqref{eqn:multiuser_multipath_mimo_model} and \eqref{eq:Y}, one can express the analog beamformer output in \eqref{eq:space_time_noise_reduced} as
\begin{equation}
\mathbf{y}^{(g)} = \left[\mathbf{X}^{(g)} \otimes \mathbf{I}_D\right]\bar{\mathbf{h}}_{eff}^{(g)} + \operatorname{vec}\left\{\left[\mathbf{S}^{(g)}\right]^H\left[\begin{array}{ccc}\bm{\eta}_0^{(g)} & \cdots & \bm{\eta}_{T-1}^{(g)}\end{array}\right]\right\},\ g=1,\ldots,G
\label{eq:y_g}
\end{equation}
where $\bm{\eta}_n^{(g)}=\sum_{g'=1,g'\neq g}^{G}\left(\sum_{k=1}^{K_{g'}}\sum_{l=0}^{L-1}\mathbf{h}_l^{(g'_k)}x_{n-l}^{(g'_k)}\right)+\mathbf{n}_n$, which is the inter-group interference to group $g$ with AWGN, and $\left\{x_n^{(g'_k)}\right\}$ for $g' \neq g$ are assumed to be composed of \textit{i.i.d.} data symbols. The true covariance matrix of $\bm{\eta}_n^{(g)}$, denoted by $\mathbf{R}_{\eta}^{(g)}$, can be calculated as $\mathbf{R}_{\eta}^{(g)}\triangleq\!\sum_{g'=1,g'\neq g}^{G}\sum_{l=0}^{L-1}\mathbf{R}_l^{(g)}$. By using the reduced dimensional observations in digital domain in \eqref{eq:y_g}, the effective channel estimates for each group can be obtained as $\hat{\bar{\mathbf{h}}}_{eff}^{(g)} = \left(\mathbf{W}^{(g)}\right)^H\mathbf{y}^{(g)}$ where $\mathbf{W}^{(g)}$ is the estimator matrix for the intended group $g$ and of size $TD_g \times K_gLD_g$. Then, the concatenated vector $\hat{\bar{\mathbf{h}}}_{eff}^{(g)}$ can be partitioned to get effective channel estimates $\hat{\mathbf{h}}_{eff,l}^{(g_k)}$ according to the structure given in \eqref{eq:h_eff_bar}.
In this stage, we utilize three different estimators in reduced dimension, described in the following subsections.

\subsection{Joint Angle-Delay Domain Reduced Rank MMSE (RR-MMSE) Estimator}
\label{sec:mmse}
By using the analog beamformer output $\mathbf{y}^{(g)}$ in \eqref{eq:y_g}, a reduced rank minimum mean square error (RR-MMSE) type estimator, adopted from \cite{Guvensen16}, can be constructed as $\hat{\bar{\mathbf{h}}}_{eff}^{(g)} \triangleq \left(\hat{\mathbf{W}}_{mmse}^{(g)}\right)^H\mathbf{y}^{(g)}$ where
\begin{gather}
\hat{\mathbf{W}}_{mmse}^{(g)}\!=\!\left(\sum_{l=0}^{L-1}\mathbf{R}^{(g)}_{code}(l)\otimes\left[\mathbf{SNR}^{(g)}_{mimo}(l)\right]\!+\!\mathbf{I}_{TD_g}\right)^{\!{-1}}\left(\sum_{l=0}^{L-1} \left(\mathbf{X}^{(g)} \left[\mathbf{I}_{K_g} \otimes \mathbf{E}_{L,l}\right]\right)\otimes \mathbf{SNR}^{(g)}_{mimo}(l)\right)
\label{eqn:W_mmse}\\
\mathbf{SNR}^{(g)}_{mimo}(l) \triangleq \left(\left[\mathbf{S}^{(g)}\right]^H \hat{\mathbf{R}}^{(g)}_{\eta}\mathbf{S}^{(g)}\right)^{-1}\left(\left[\mathbf{S}^{(g)}\right]^H \hat{\mathbf{R}}_l^{(g)}\mathbf{S}^{(g)}\right)
\label{eqn:snr_mimo_l} \\
\mathbf{R}^{(g)}_{code}(l) \triangleq \left(\mathbf{X}^{(g)}\left[\mathbf{I}_{K_g} \otimes \mathbf{E}_{L,l}\right]\left[\mathbf{X}^{(g)}\right]^H \right).
\label{eqn:r_code_l}
\end{gather}
\noindent It is an approximated MMSE estimator where the sparsity map in \eqref{eq:sparsity_map} and estimated CCMs in \eqref{eq:est_ccm_def} are utilized instead of their true values in the formulation\footnote{
While deriving \eqref{eqn:W_mmse}, it is assumed that each user in the same group has \textit{i.i.d.} channels with $\mathcal{CN}\left(\mathbf{0},\mathbf{R}_l^{(g)}\right)$ where $\mathbf{R}_l^{(g)}\triangleq\sum_{k=1}^{K_g}\mathbf{R}_l^{(g_k)}$ for $g=1,\ldots,G$. This is a common assumption for JSDM framework \cite{adhikary14}, \cite{nam14}, \cite{Guvensen16}.}. 
In (\ref{eqn:W_mmse}), $\mathbf{E}_{L,l}$ is an $L \times L$ elementary diagonal matrix where all the entries except the $\left(l+1\right)^{th}$ diagonal one are zero. In (\ref{eqn:snr_mimo_l}), $\hat{\mathbf{R}}_\eta^{(g)}$ is the estimated correlation matrix of inter-group interference to group $g$ given in \eqref{eqn:group_interference_ccm}. If $\hat{\mathbf{R}}_l^{(g)}$ is a perfect estimate, the estimator in \eqref{eqn:W_mmse} becomes the optimal reduced rank linear estimator (Wiener filter) in MSE sense \cite{guvensen16_2}, whose performance is to be used as benchmark.

\subsection{Joint Angle-Delay Domain Beamspace Aware Least Square (BA-LS) Estimator}
\label{sec:ba_ls}
One can simplify the RR-MMSE estimator in \eqref{eqn:W_mmse} by assuming that the eigenspaces of each MPCs after analog beamforming are almost orthogonal, an effect more strongly observed in mm-wave channels especially for the case of large number of antenna elements \cite{guvensen16_2}. Based on this orthogonality assumption, we can construct an estimator, called as BA-LS, which exploits only the joint angle-delay sparsity of the channels, as $\hat{\bar{\mathbf{h}}}_{eff}^{(g)} \triangleq \left(\mathbf{W}_{ba-ls}^{(g)}\right)^H\mathbf{y}^{(g)}$ where
\begin{equation}
\mathbf{W}_{ba-ls}^{(g)} = \sum_{\ell=0}^{\mathcal{MPC}^{(g)}-1}
\left( \operatorname{pinv}\left\{\left[\mathbf{I}_{K_g} \otimes \sum_{m \in \mathfrak{L}_\ell^{(g)}} \mathbf{E}_{L,m}\right]
\left[\mathbf{X}^{(g)}\right]^H\right\} \otimes \left[\sum_{n \in \mathfrak{D}_\ell^{(g)}}\mathbf{E}_{D_g,n}\right]\right).
\label{eqn:W_bals}
\end{equation}
In (\ref{eqn:W_bals}), $\mathcal{MPC}^{(g)}$ is the total number of MPC clusters, resolvable in angular domain, in group $g$ having nearly non-overlapping AoA support. The set $\mathfrak{L}_\ell^{(g)}$ in \eqref{eq:resolvable_mpc_groups}, which is obtained via the estimated sparsity map $\mathbf{I}^{(k)}$'s, is composed of estimated active (temporal) delays belonging to the $\ell^{th}$ resolvable multipath group having MPCs with significantly overlapping AoA support in the angular domain for $\ell=1,\ldots,\mathcal{MPC}^{(g)}-1$. In (\ref{eqn:W_bals}), the set $\mathfrak{D}_\ell^{(g)}$ is defined as $\mathfrak{D}_\ell^{(g)} \triangleq \left\{ n \in \mathbb{Z}^+ \vert \sum_{m=0}^{\ell-1}d_m^{(g)} < n \leq \sum_{m=0}^{\ell}d_m^{(g)} \right\}$ for $\ell=1,\ldots,\mathcal{MPC}^{(g)}-1$, and $\mathfrak{D}_\ell^{(g)} \triangleq \left\{ n \in \mathbb{Z}^+ \vert 0 < n \leq d_0^{(g)} \right\}$ for $\ell=0$ where $\vert\mathfrak{D}_\ell^{(g)}\vert=d_\ell^{(g)}$. Here, $\mathfrak{D}_\ell^{(g)}$ shows the column indices of the analog beamformer matrix $\mathbf{S}^{(g)}$ in Section \ref{sec:abeam_design} allowing to pass the $\ell^{th}$ resolvable MPC cluster of group $g$. For BA-LS estimator, only the estimated sparsity map $\mathbf{I}^{(k)}$ in \eqref{eq:sparsity_map} is necessary to construct $\mathfrak{L}_{\ell}^{(g)}$. Here, $\operatorname{pinv}\left\{\;\right\}$, denoting the pseudo-inverse operation, can be seen as the temporal correlator preceded by the analog beamformer. It performs the task of LS type estimation of reduced dimensional channels corresponding to the $\ell^{th}$ MPC cluster in group $g$. Furthermore, with the help of $\mathbf{S}^{(g)}$, which orthogonalize the MPCs in reduced dimensional beamspace (eigenspace), the estimation of channel response for each MPCs can be fulfilled separately in their individual beamspaces. This reduces $T_{fast}$ significantly compared to conventional LS type estimators.

\subsection{Conventional LS Estimator}
\label{sec:ls}
\begin{figure*}[!t]
\centering
\includegraphics[width=0.95\linewidth]{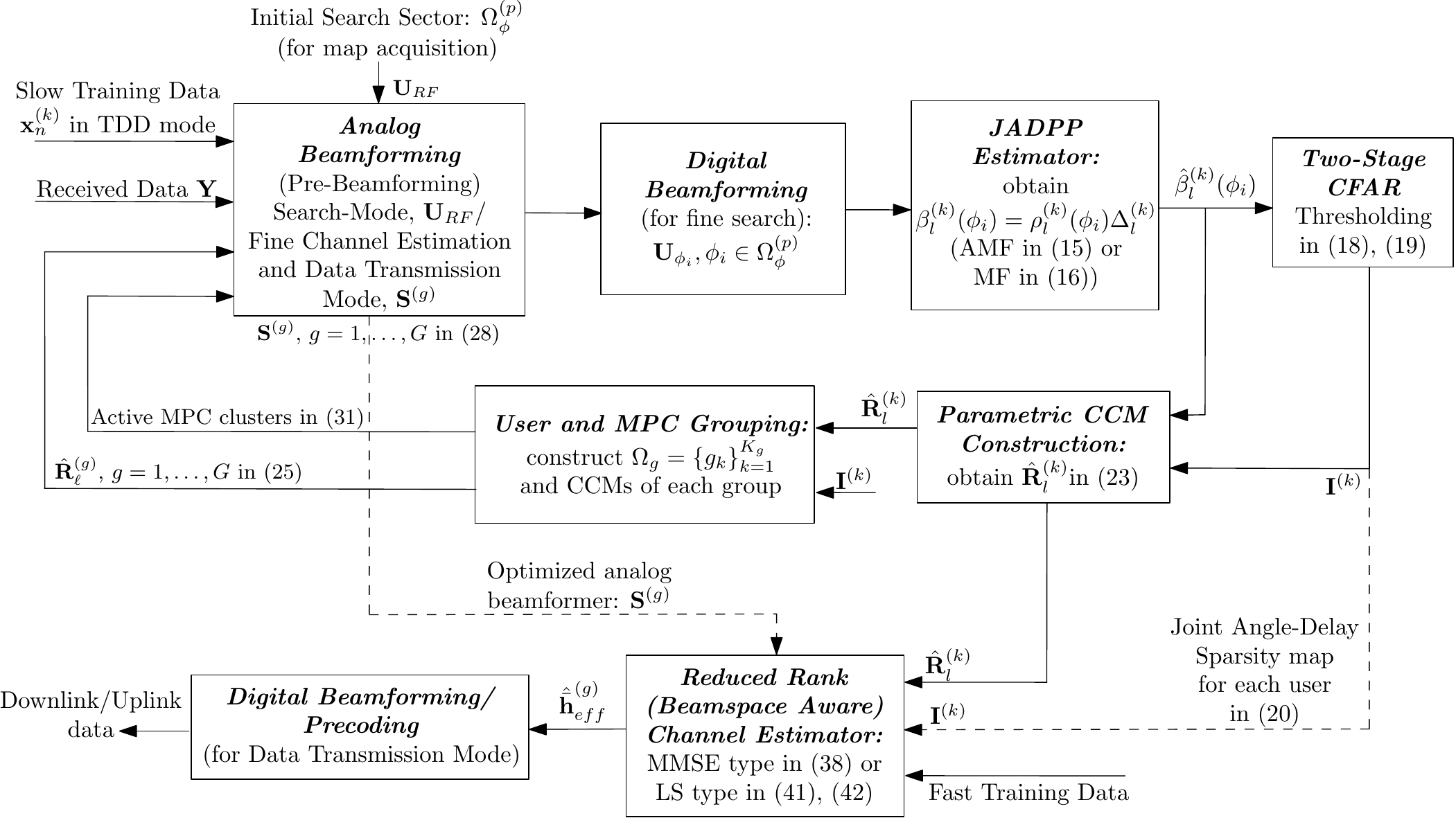}
\caption{Overall System Architecture}
\label{fig:overall_model}  
\end{figure*}

Different from \eqref{eqn:W_bals}, LS type estimator can be constructed in a conventional way \cite{ashikhmin11}, after analog beamforming, as $\hat{\bar{\mathbf{h}}}_{eff}^{(g)} \triangleq \left(\mathbf{W}_{ls}^{(g)}\right)^H\mathbf{y}^{(g)}$ where
\begin{equation}
\mathbf{W}_{ls}^{(g)} = 
\left\{ \begin{array}{cc}
\left[\mathbf{X}^{(g)}\right]\left(\left[\mathbf{X}^{(g)}\right]^H\mathbf{X}^{(g)}\right)^{-1} \otimes 
\left[\mathbf{I}_{D_g}\right] &
\qquad \textrm{ if } T_{fast} \geq K_gL \\[5pt]
\left(\mathbf{X}^{(g)}\left[\mathbf{X}^{(g)}\right]^H\right)^{-1}
\left[\mathbf{X}^{(g)}\right] \otimes 
\left[\mathbf{I}_{D_g}\right] &
\qquad \textrm{ if } T_{fast} < K_gL
\end{array},\right.
\label{eqn:W_ls}
\end{equation}
and $\mathbf{X}^{(g)}$ in (\ref{eqn:complete_training_matrix}) is assumed to be full column or row rank. In \eqref{eqn:W_ls}, a conventional LS estimator in reduced dimensional beamspace (formed by $\mathbf{S}^{(g)}$) is constructed without taking the joint angle-delay sparsity of the channels into account.\\
\indent Finally, the overall hybrid beamforming based system architecture consisting of initial slow-time beam acquisition mode (JADPP, Two-Stage CFAR thresholding, CCM and sparsity map construction together with analog beamformer update), and fast-time fine channel acquisition mode (RR-MMSE, BA-LS, LS in reduced beamspace) are summarized in Fig. \ref{fig:overall_model}.  

\section{Performance Evaluation}
\label{sec:mse}
In order to compare the performances of proposed JADPP estimators, namely AMF in \eqref{eq:amf_est} and MF in \eqref{eq:mf_est} and, different channel estimators, namely MMSE, BA-LS and conventional LS in \eqref{eqn:W_mmse}, \eqref{eqn:W_bals} and \eqref{eqn:W_ls}, we define the following performance metric below
\begin{equation}
nMSE^{(g)}=\frac{\mathbb{E}_{\mathbf{W}^{(g)},\mathbf{S}^{(g)}}\left\{\mathbb{E}\left\{||\bar{\mathbf{h}}_{eff}^{(g)}-\hat{\bar{\mathbf{h}}}_{eff}^{(g)}||^2\ |\ \mathbf{W}^{(g)},\mathbf{S}^{(g)}\right\}\right\}}{\mathbb{E}_{\mathbf{S}^{(g)}}\left\{\mathbb{E}\left\{||\bar{\mathbf{h}}_{eff}^{(g)}||^2\ |\ \mathbf{S}^{(g)}\right\}\right\}}
\label{eqn:nmse}
\end{equation}
which shows the normalized MSE (nMSE) of the channel estimates for group $g$ users. In \eqref{eqn:nmse}, $\hat{\bar{\mathbf{h}}}_{eff}^{(g)} = \left(\mathbf{W}^{(g)}\right)^H\mathbf{y}^{(g)}$, where $\mathbf{W}^{(g)}$ can be seen as an arbitrary linear estimator. For performance evaluation, $\hat{\mathbf{W}}_{mmse}^{(g)}$, $\mathbf{W}_{ba-ls}^{(g)}$, $\mathbf{W}_{ls}^{(g)}$ are to be used in place of $\mathbf{W}^{(g)}$ while CCMs are constructed by AMF or MF type preprocessing (in pre-grouping stage). Given $\mathbf{S}^{(g)}$ and $\mathbf{W}^{(g)}$, the numerator and denominator of the conditional expectations in \eqref{eqn:nmse} can be found analytically as\footnote{
Here, Block Fading channel model is assumed where each user channel changes independently from block to block. Therefore, the expectations in \eqref{eqn:nmse} can be found by using Monte Carlo (MC) based averaging over independent training snapshots where independent JADPP estimates and CCMs are obtained to construct $\mathbf{S}^{(g)}$ and $\mathbf{W}^{(g)}$.}
\begin{align}
\mathbb{E}\left\{||\bar{\mathbf{h}}_{eff}^{(g)}\scalebox{0.8}[1.0]{\( - \)}\hat{\bar{\mathbf{h}}}_{eff}^{(g)}||^2\ |\ \mathbf{W}^{(g)}, \mathbf{S}^{(g)}\right\} &= \mathbb{E}\left\{\operatorname{Tr}\left\{{\left(\bar{\mathbf{h}}_{eff}^{(g)}\ \scalebox{0.8}[1.0]{\( - \)}\left(\mathbf{W}^{(g)}\right)^H\mathbf{y}^{(g)}\right)}{\left(\bar{\mathbf{h}}_{eff}^{(g)}\ \scalebox{0.8}[1.0]{\( - \)}\left(\mathbf{W}^{(g)}\right)^H\mathbf{y}^{(g)}\right)}^H\right\}\right\} \nonumber \\
&= \operatorname{Tr}\left\{\mathbf{R}_{mmse}^{(g)}\right\}+\operatorname{Tr}\left\{\left(\mathbf{W}^{(g)}\scalebox{0.8}[1.0]{\( - \)}\mathbf{W}_{mmse}^{(g)}\right)^H\mathbf{R}_{y}^{(g)}\left(\mathbf{W}^{(g)}\scalebox{0.8}[1.0]{\( - \)}\mathbf{W}_{mmse}^{(g)}\right)\right\} \label{eq:nmse_num}\\
\mathbb{E}\left\{||\bar{\mathbf{h}}_{eff}^{(g)}||^2\ |\ \mathbf{S}^{(g)}\right\} &= \operatorname{Tr}\left\{\mathbf{R}_{eff}^{(g)}\right\}
\end{align}
where
\begin{align}
\mathbf{R}_{eff}^{(g)} &\triangleq \mathbb{E}\left\{\bar{\mathbf{h}}_{eff}^{(g)}\left[\bar{\mathbf{h}}_{eff}^{(g)}\right]^H\right\} = \sum_{l=0}^{L-1}\left[\mathbf{I}_{K_{g}}\otimes\mathbf{E}_{L,l}\right]\otimes\left(\left[\mathbf{S}^{(g)}\right]^H\mathbf{R}_{l}^{(g)}\left[\mathbf{S}^{(g)}\right]\right)\nonumber \\
\mathbf{R}_{mmse}^{(g)} &\triangleq \mathbb{E}\left\{\mathbf{e}^{(g)}\left(\mathbf{e}^{(g)}\right)^H\ \!|\ \!\mathbf{W}^{(g)}=\mathbf{W}_{mmse}^{(g)}\right\} = \mathbf{R}_{eff}^{(g)}-\mathbf{A}\left(\mathbf{R}_{y}^{(g)}\right)^{\!{-1}}\mathbf{A}^H,\ \ \mathbf{e}^{(g)} \triangleq \bar{\mathbf{h}}_{eff}^{(g)}-\left[\mathbf{W}_{mmse}^{(g)}\right]^H\mathbf{y}^{(g)} \nonumber \\
\mathbf{R}_{y}^{(g)} &\triangleq \mathbb{E}\left\{\mathbf{y}^{(g)}(\mathbf{y}^{(g)})^H\right\} \nonumber \\
&= \sum_{l=0}^{L-1}\left(\mathbf{X}^{(g)}\left(\mathbf{I}_{K_{g}}\otimes\mathbf{E}_{L,l}\right)\left(\mathbf{X}^{(g)}\right)^H\right)\otimes\left(\left(\mathbf{S}^{(g)}\right)^H\mathbf{R}_{l}^{(g)}\left(\mathbf{S}^{(g)}\right)\right)+\mathbf{I}_T\otimes\left(\left(\mathbf{S}^{(g)}\right)^H\mathbf{R}_{\eta}^{(g)}\left(\mathbf{S}^{(g)}\right)\right) \nonumber \\
\mathbf{A}&=\sum_{l=0}^{L-1}\left(\mathbf{I}_{K_{g}}\otimes\mathbf{E}_{L,l}\right)\left(\mathbf{X}^{(g)}\right)^H\otimes\left(\left[\mathbf{S}^{(g)}\right]^H\mathbf{R}_{l}^{(g)}\left[\mathbf{S}^{(g)}\right]\right).
\end{align}
In \eqref{eq:nmse_num}, $\mathbf{R}_{mmse}^{(g)}$ is the MMSE covariance matrix for the case when optimal MMSE estimator called as $\mathbf{W}_{mmse}^{(g)}$, with perfect knowledge of CCMs are used in place of $\mathbf{W}^{(g)}$. 

\section{Numerical Results and Discussion}
\label{sec:num_results}
\begin{figure}[tb]
\includegraphics[width=0.5\linewidth]{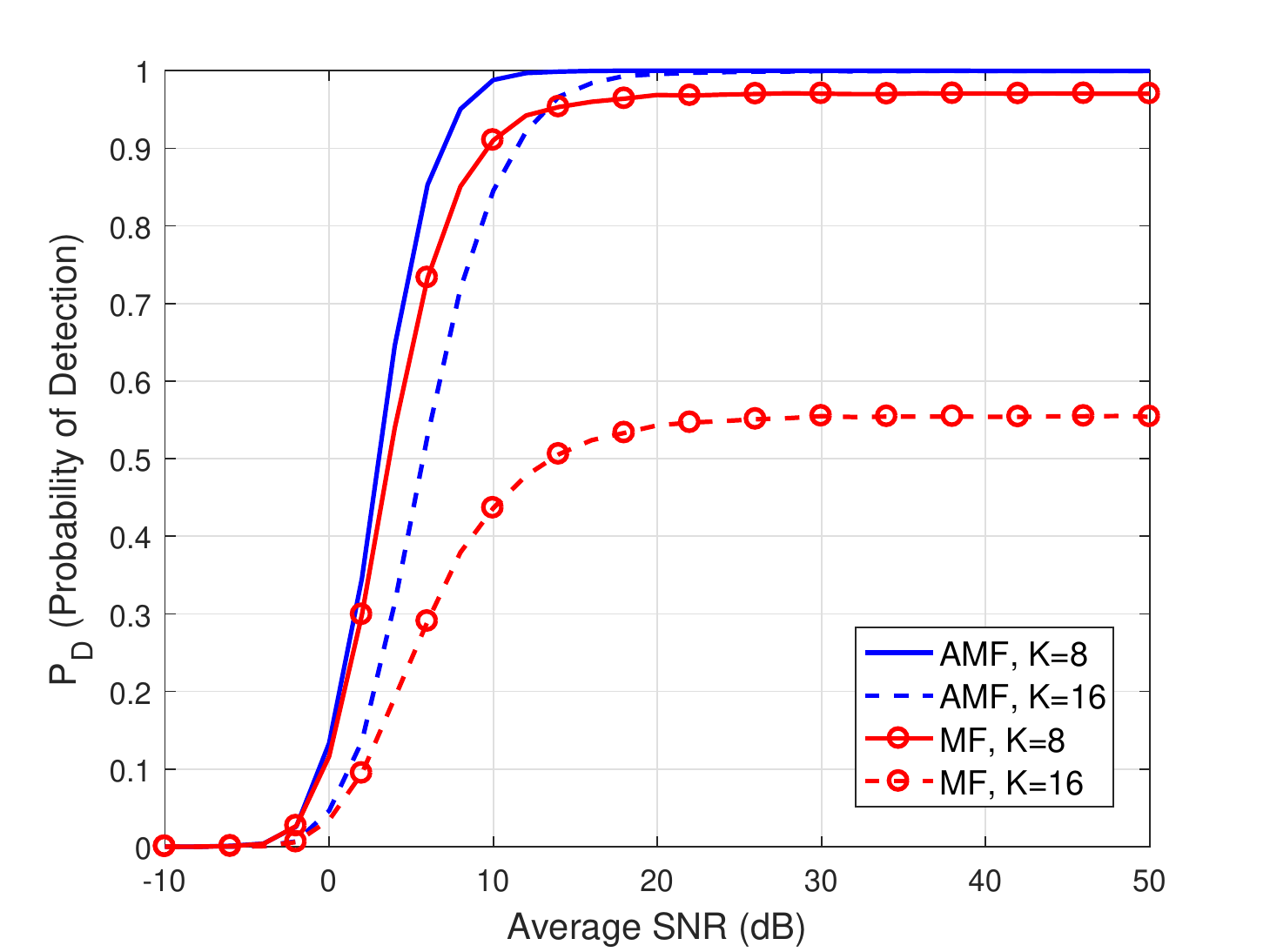}\hfill
\includegraphics[width=0.5\linewidth]{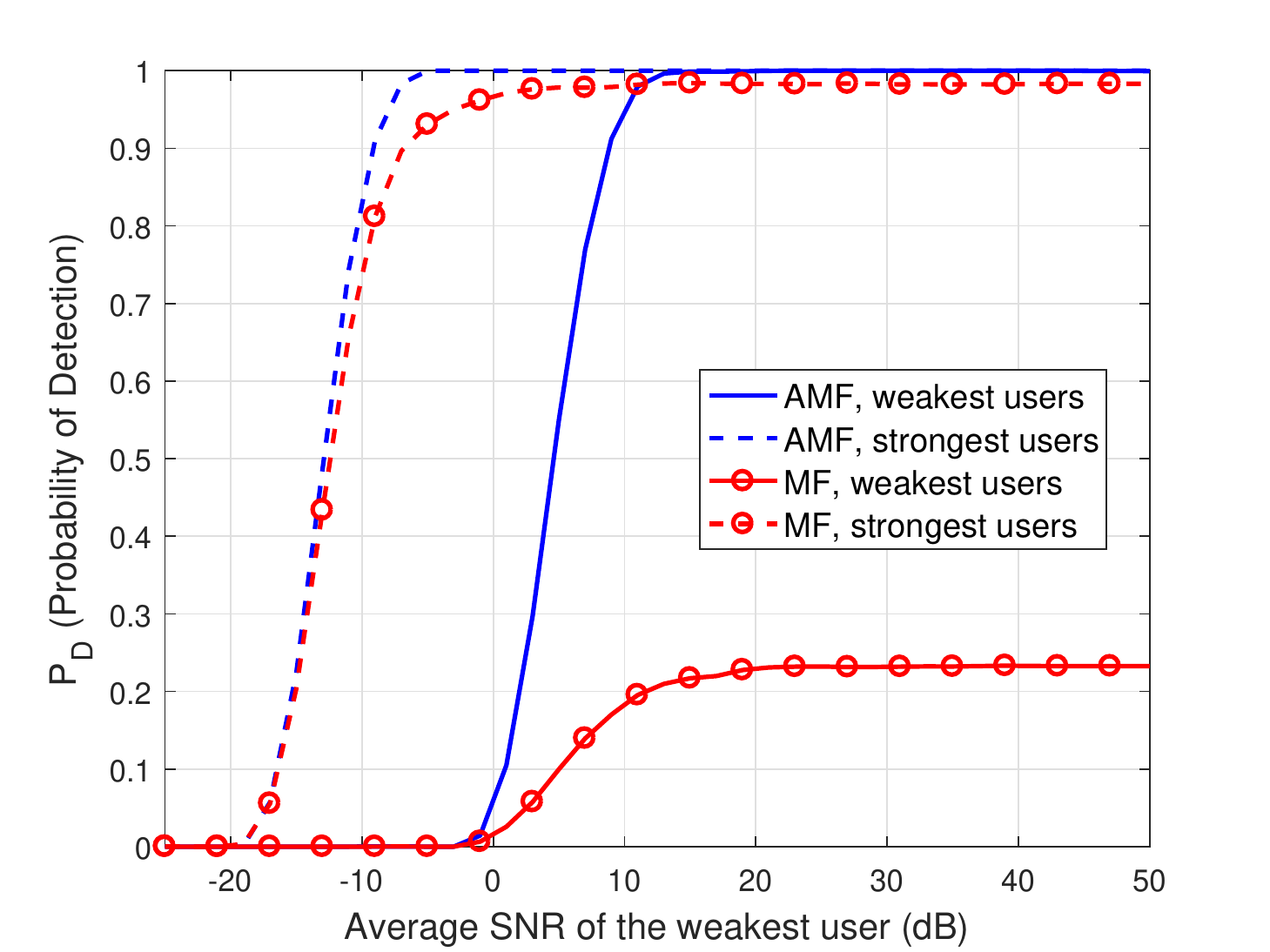}
{\phantomsubcaption\label{fig:same_pd}}
{\phantomsubcaption\label{fig:diff_pd}}
\caption{Average $P_D$ vs SNR curves for equal (left) and different (right) averaged received power levels ($T=32$, $J=5$)}
\label{fig:pd}
\end{figure}

\begin{figure}[tb]
\includegraphics[width=0.5\linewidth]{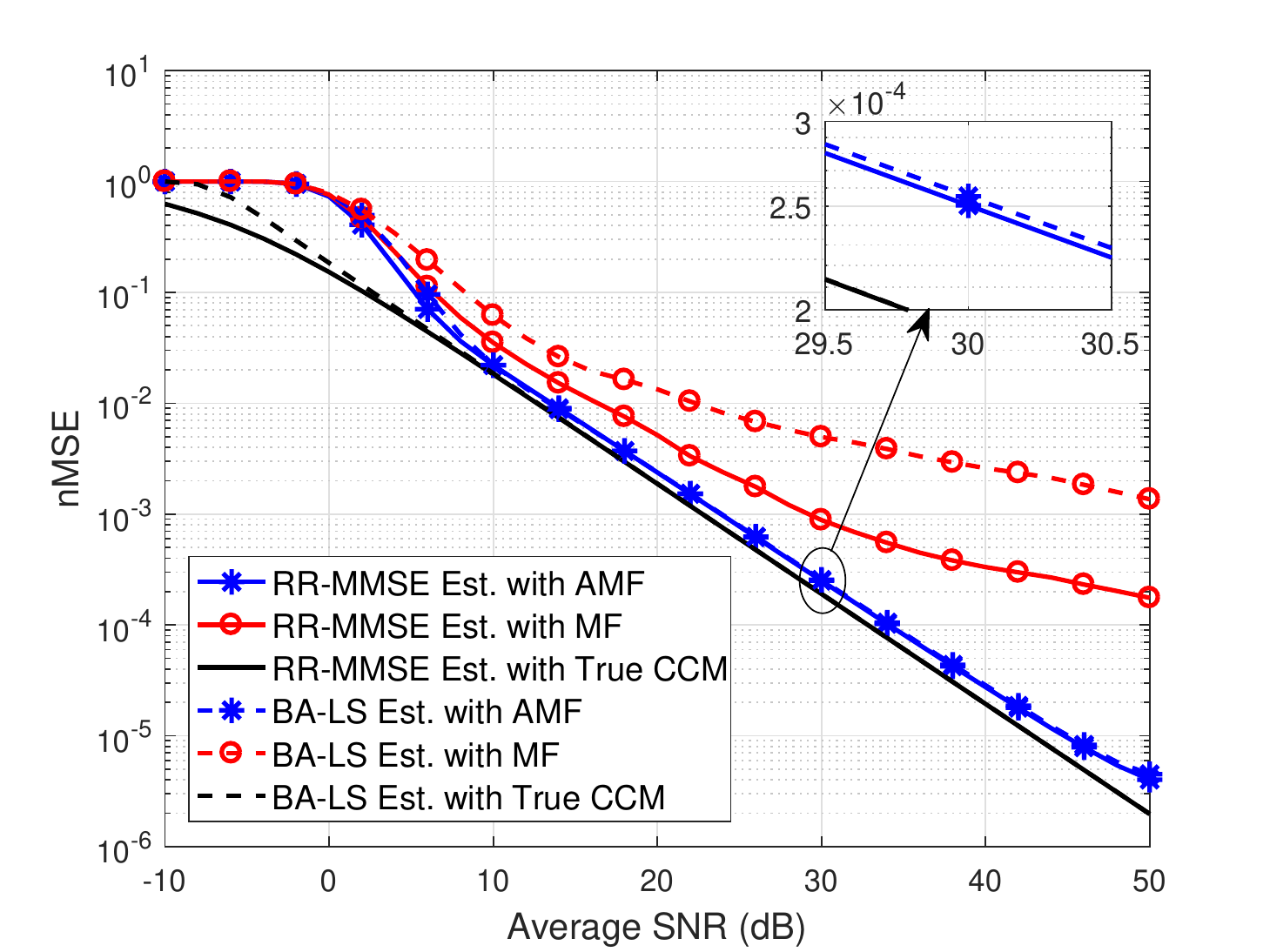}\hfill
\includegraphics[width=0.5\linewidth]{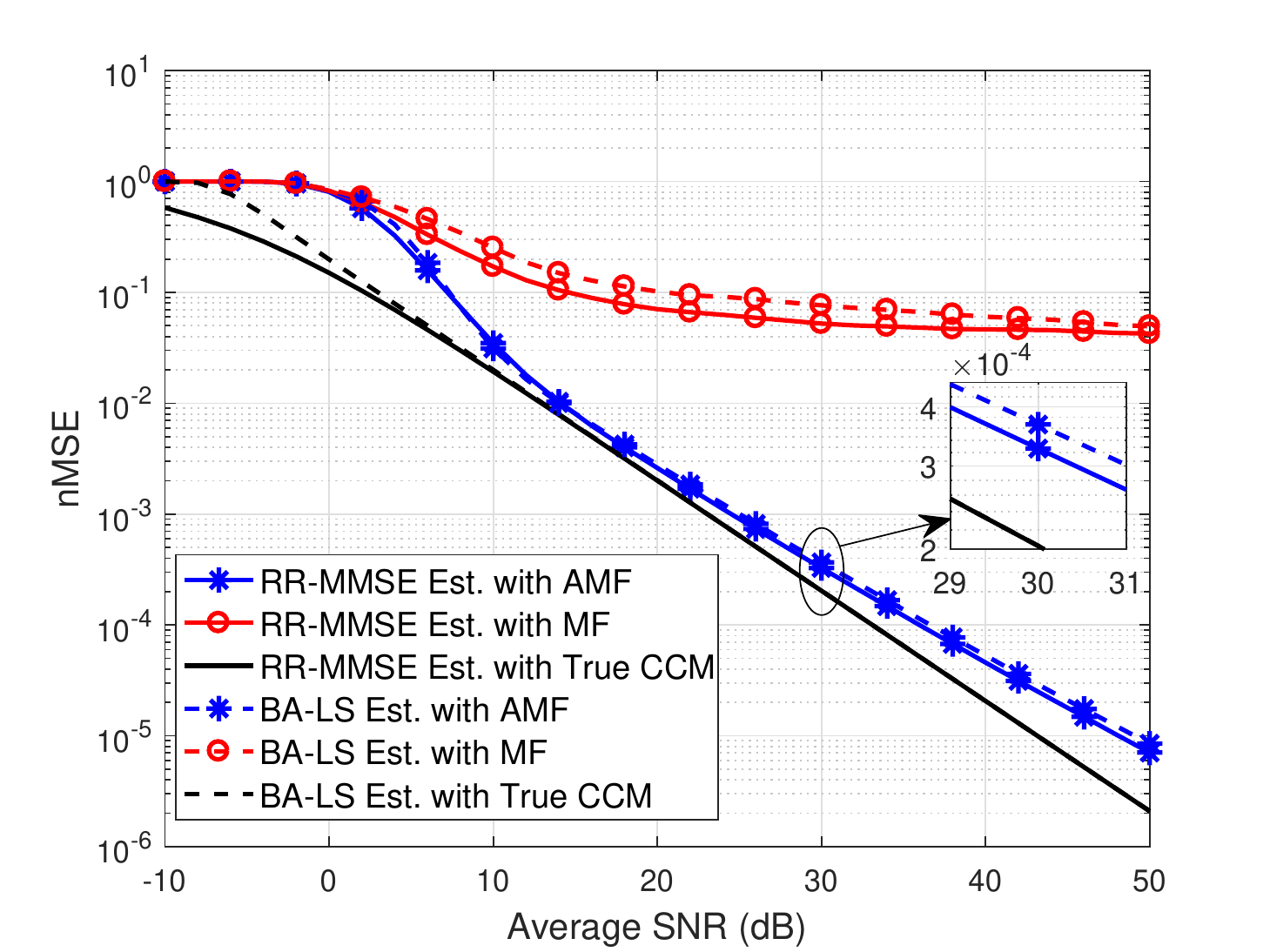}
{\phantomsubcaption\label{fig:s10_same_mse}}
{\phantomsubcaption\label{fig:s11_same_mse}}
\caption{Performance of different channel estimators averaged over all groups for $K=8$ (left) and $K=16$ (right) cases ($T=T_{fast}=32$, $J=5$)}
\label{fig:same_mse}
\end{figure}

\begin{figure}[tb]
\includegraphics[width=0.5\linewidth]{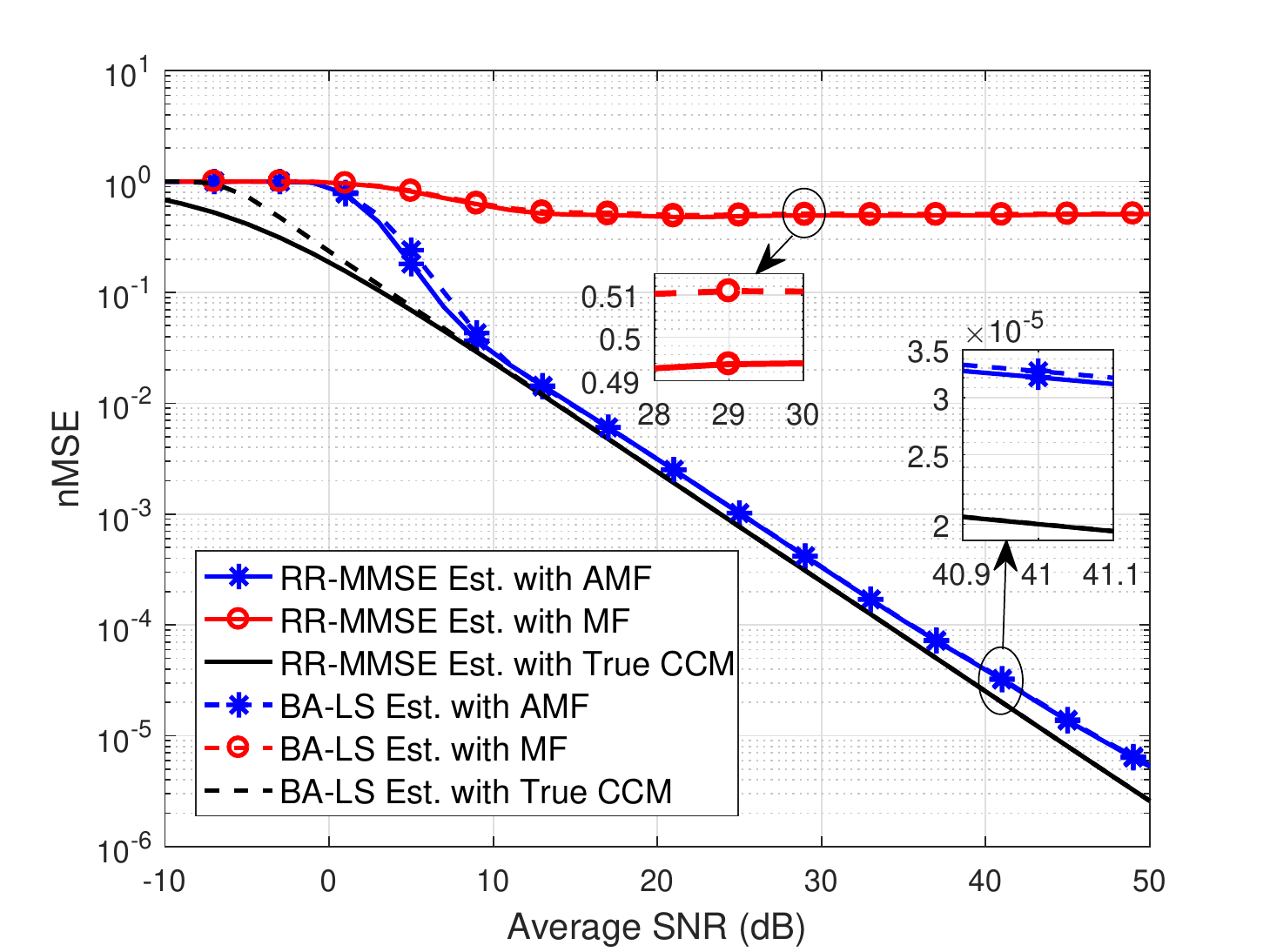}\hfill
\includegraphics[width=0.5\linewidth]{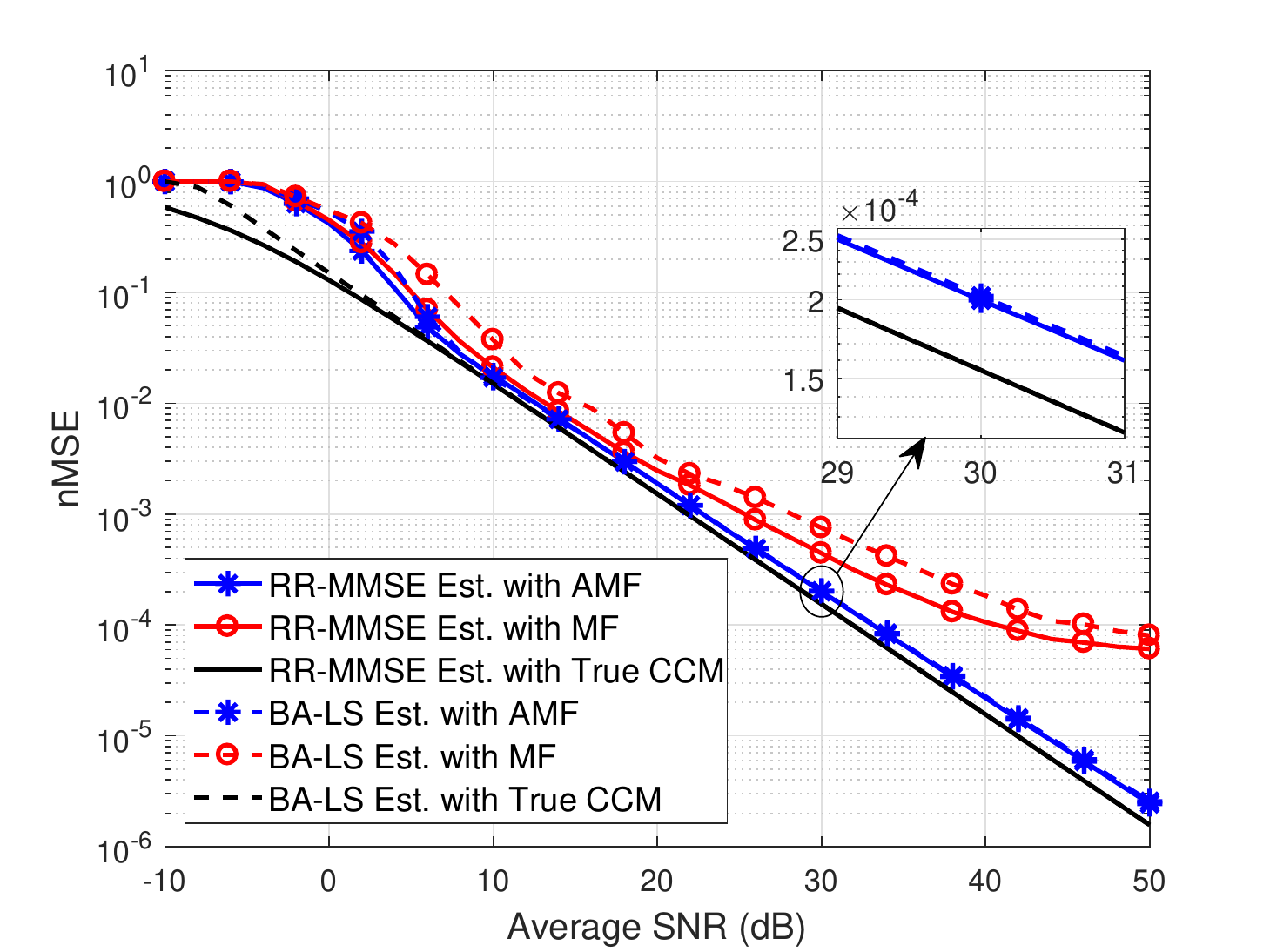}
{\phantomsubcaption\label{fig:s10_diff_mse_weak}}
{\phantomsubcaption\label{fig:s10_diff_mse_strong}}
\caption{Performance of different channel estimators for weakest (left) and strongest (right) user groups ($K=8$, $T=T_{fast}=32$, $J=5$)}
\label{fig:diff_mse}
\end{figure}

\begin{figure}[tb]
\includegraphics[width=0.5\linewidth]{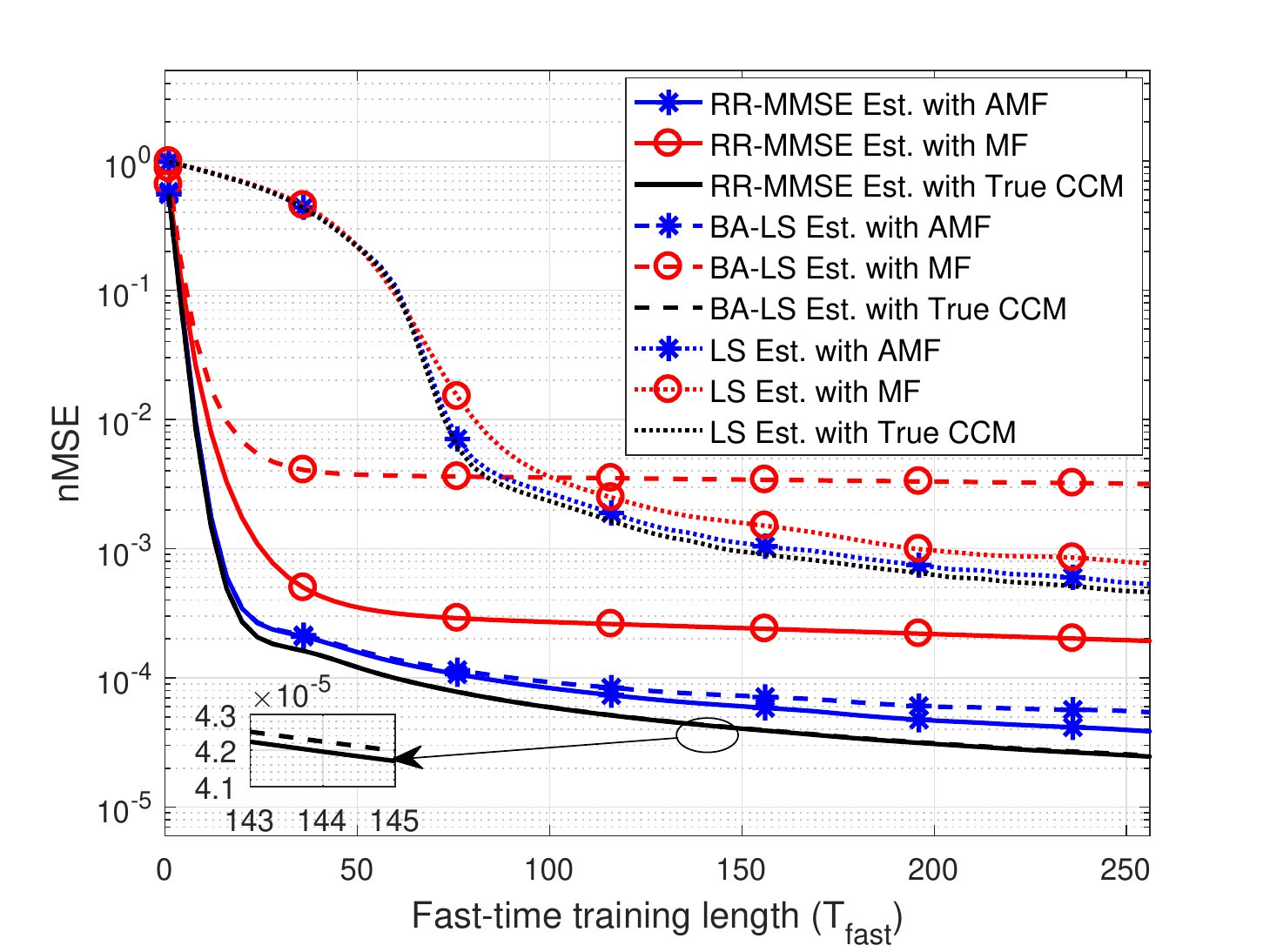}\hfill
\includegraphics[width=0.5\linewidth]{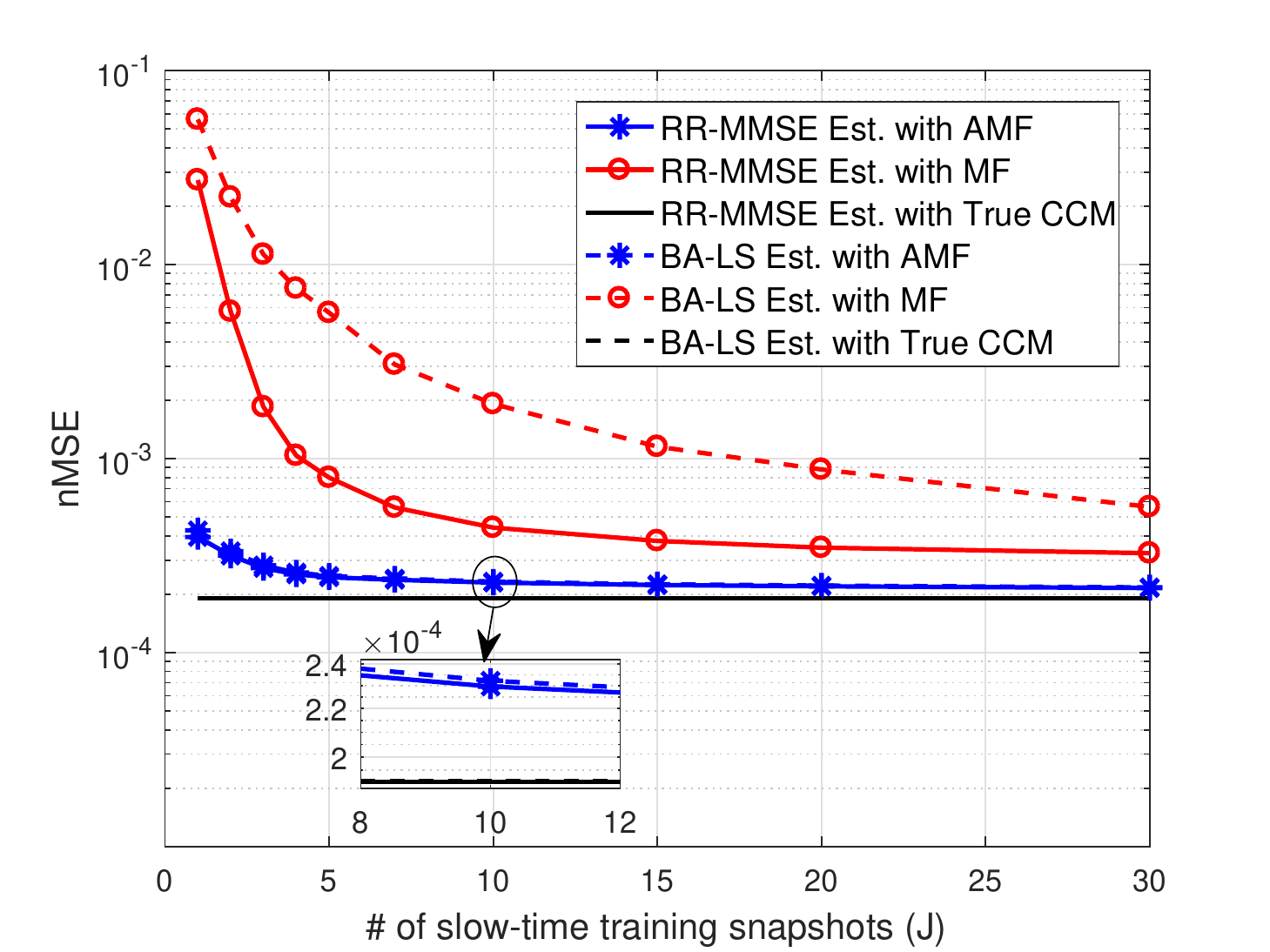}
{\phantomsubcaption\label{fig:s10_same_T_30dB}}
{\phantomsubcaption\label{fig:s10_same_snapshot}}
\caption{Performance of different channel estimators averaged over all groups in terms of nMSE vs $T_{fast}$ (left) and $J$ (right) ($K=8$, $\text{SNR}^{(g)}=30$ dB for $g=1,\ldots,G$)}
\label{fig:T_J}
\end{figure}

Here, we provide some numerical results to evaluate the performance of AMF and MF type JADPP estimators (given in Section \ref{sec:sec_jadpp}) via probability of detection ($P_D$) curves. Also, performance of the channel estimators (given in Section \ref{sec:chan_est}) with respect to SNR, fast-time training sequence length ($T_{fast}$) and number of slow-time training snapshots ($J$) are demonstrated for different number of users ($K$) via nMSE curves. Throughout the demonstrations, we consider a massive MIMO system with uplink training in TDD mode where the BS is equipped with a ULA of $N=100$ antenna elements, and each of $K$ users has a single antenna. It is assumed that $K=8$ or $16$ users were clustered into four groups ($G=4$) where the true angle-delay profile of all UEs for $K=16$ case is shown in Fig. \ref{fig:user_locations}, while numerical values of mean AoA and AS of each MPCs are provided in Table \ref{table:user_locations}. If $K$ is taken as $8$, two users to be active were selected randomly among others in each group\footnote{
Here, we assume that users come in groups, either by nature or by the use of proper user grouping algorithms.
}. Due to hybrid beamforming adoption, limited number of RF chains are utilized, and equal number of RF chains are assigned to each group, then $D_g=D/G$. In beam acquisition mode, the angular search sector of interest $\Omega_\phi$ is determined to be $(-45\degree,45\degree)$. While constructing the JADPP and sparsity map of all active users, the angular resolution in $\phi$ is taken as $0.25\degree$ ($M=\frac{90\degree}{0.25\degree}=360$) to construct fine digital beams, number of RF chains $D$ is set to $8$, and the length of the slow-time training data $T$ in \eqref{eq:xlk} is taken to be equal to the channel memory $L=32$. For AMF in \eqref{eq:amf_est} and MF in \eqref{eq:mf_est}, the number of digital beams $D_{search}$ is set to $5$, look spread $\sigma$ in \eqref{eq:Rphi} is taken as $3\degree$, and $J$ is taken as $5$. In CFAR thresholding, where the sparsity map of all active users are obtained, the desired false alarm rate $\bar{P}_{FA}$ is set to $10^{-3}$ in \eqref{eq:test_in_temp} and \eqref{eq:test_in_space}, and the length of guard interval $\Pi_{\phi_i}$ in \eqref{eq:test_in_space} is taken as $4\degree$. In our scenario, simultaneously active users use non-orthogonal training waveforms composed of $L=32$ chips, and these are obtained by truncating length-63 \textit{Kasami} codes without any optimization\footnote{
There are more efficient approaches (other than the truncation of Kasami codes) yielding waveforms with better cross- and auto correlation properties, but training optimization is beyond the scope of this work.
}. In our simulation setup, at the BS, the average received signal strength of different UEs in the same group are assumed to be equal. Moreover, in order to observe the near-far effect\footnote{
The near-far effect stems from the fact that the average received signal strength of different UEs may differ significantly depending on their distance to the BS.
}, at the BS, the case of unequal average received power of UEs in different groups is investigated. The average received group power is defined as $\bar{\beta}^{(g)}\triangleq\frac{1}{K_g}\sum_{k=1}^{K_g}\beta^{(g_k)}$. Two different cases are considered. In the first case $\bar{\beta}^{(g)}$'s are assumed to be equal, and in the second case, different average received power levels at BS are assumed for different groups, in which $10\log\left(\frac{\bar{\beta}^{(4)}}{\bar{\beta}^{(3)}}\right)=10\log\left(\frac{\bar{\beta}^{(3)}}{\bar{\beta}^{(2)}}\right)=10\log\left(\frac{\bar{\beta}^{(2)}}{\bar{\beta}^{(1)}}\right)=5$ dB are taken. While constructing the $P_D$ in \eqref{eq:pd_def}, and nMSE in \eqref{eqn:nmse}, Monte Carlo based averaging technique is utilized by taking sufficiently high number of realizations.\\
\begin{table}[tb]
  \begin{varwidth}[b]{0.5\linewidth}
	\centering
	\resizebox{0.85\textwidth}{!}{
	\begin{tabular}{V{2.5}c|c|c|c|cV{2.5}}
  \Xhline{1pt}
	$k$ & $g$ & \makecell{Active MPC \\index} & \makecell{Mean \\AoA} & \makecell{Angular \\Spread} \\
	\Xhline{1pt}
  1 & \multirow{4}{*}{1} & \{0, 5, 11\} & \{0, 9.75, 22\} & \{3, 2.5, 3.5\} \\
	\cline{1-1} \cline{3-5}
  2 & &\{1, 4, 13\} & \{-0.5, 8.5, 20.75\} & \{3, 2, 2.5\} \\
	\cline{1-1} \cline{3-5}
	3 & &\{0, 4, 12\} & \{1, 9.25, 21\} & \{3, 2.5, 3.5\} \\
	\cline{1-1} \cline{3-5}
	4 & &\{1, 5, 13\} & \{1.5, 9, 21.5\} & \{3, 2, 2.5\} \\
	\Xhline{1pt}
	5 & \multirow{4}{*}{2} & \{3, 9\} & \{27.5 15.25\} & \{3, 2\} \\
	\cline{1-1} \cline{3-5}
	6 & & \{2, 12\} & \{26.5 14.75\} & \{3.5, 2.5\} \\
	\cline{1-1} \cline{3-5}
	7 & & \{3, 10\} & \{26.5 15.75\} & \{3, 2\} \\
	\cline{1-1} \cline{3-5}
	8 & & \{2, 11\} & \{27.15 15\} & \{3.5, 2.5\} \\
	\Xhline{1pt}
	9 & \multirow{4}{*}{3} & \{8, 17\} & \{-6.25 -13.5\} & \{3.5, 3\} \\
	\cline{1-1} \cline{3-5}
	10 & & \{10, 19\} & \{-8 -14.25\} & \{3.5, 3\} \\
	\cline{1-1} \cline{3-5}
	11 & & \{6, 18\} & \{-6.75 -14\} & \{3.5, 3\} \\
	\cline{1-1} \cline{3-5}
	12 & & \{7, 19\} & \{-7.5 -13.5\} & \{3, 3.5\} \\
	\Xhline{1pt}
	13 & \multirow{4}{*}{4} & \{20, 29\} & \{-20.5 -28\} & \{3, 3.5\} \\
	\cline{1-1} \cline{3-5}
	14 & & \{18, 25\} & \{-19.75 -27\} & \{2 2.5\} \\
	\cline{1-1} \cline{3-5}
	15 & & \{21, 29\} & \{-20 -27.25\} & \{3, 3.5\} \\
	\cline{1-1} \cline{3-5}
	16 & & \{19, 26\} & \{-20.25 -27\} & \{2, 2.5\} \\
	\Xhline{1pt}
\end{tabular}}
	\captionsetup{justification=centering}
    \caption{True angle-delay profile of all users}
    \label{table:user_locations}
  \end{varwidth}%
  \hfill
  \begin{minipage}[b]{0.5\linewidth}
    \centering
		\includegraphics[width=0.9\linewidth]{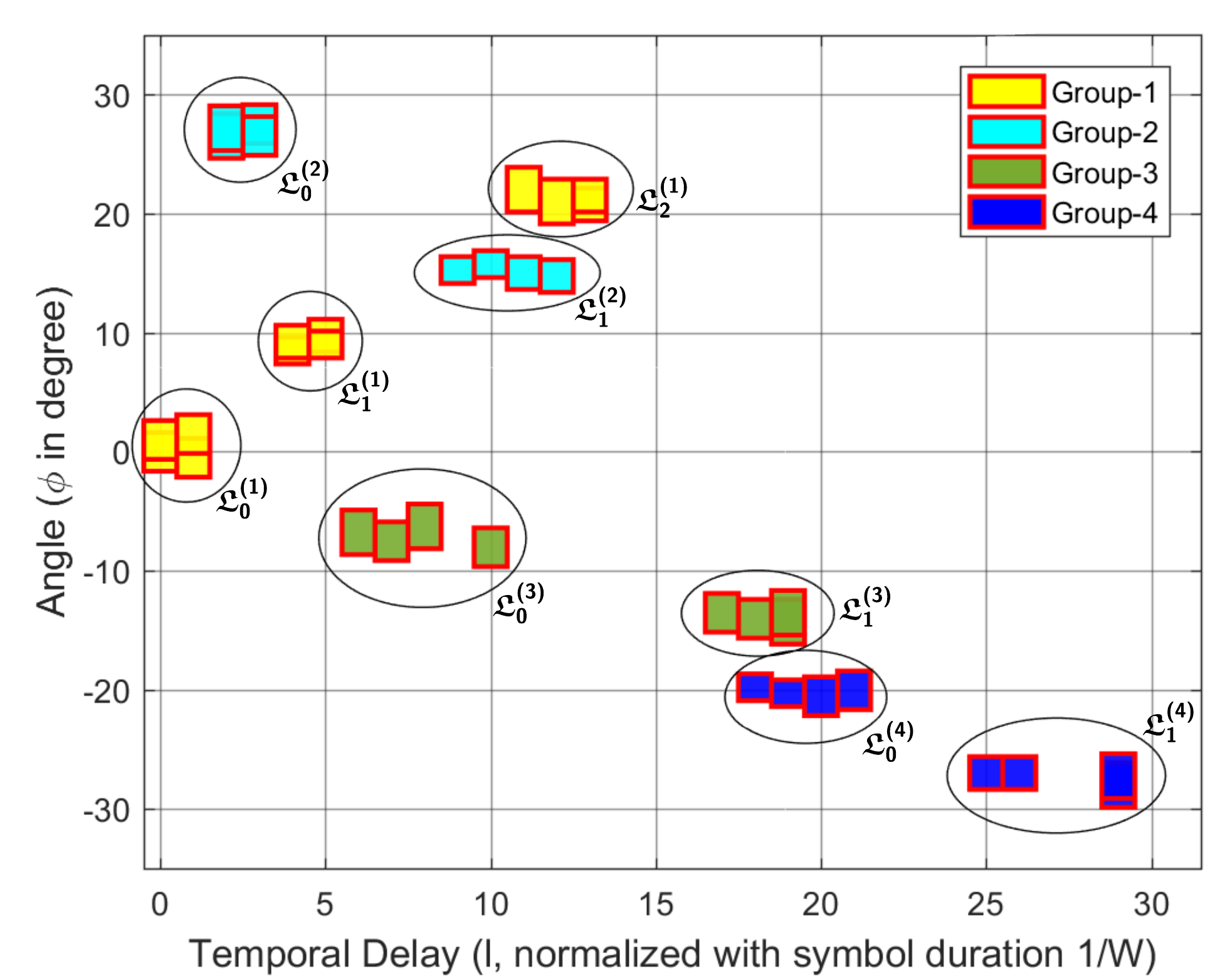}
    \captionof{figure}{Joint angle-delay map of all users}
    \label{fig:user_locations}
  \end{minipage}
\end{table}
\indent In Fig. \ref{fig:same_pd} and Fig. \ref{fig:diff_pd}, performances of CFAR thresholding based on AMF and MF type JADPP estimators in terms of average $P_D$ with respect to average group SNR, namely $\text{SNR}^{(g)}=\frac{\bar{\beta}^{(g)}}{N_0}$ are given. In Fig. \ref{fig:same_pd}, when $\bar{\beta}^{(g)}$'s are equal, AMF outperforms MF for $8$ and $16$ user cases. The difference between two methods is more apparent when the number of simultaneously active users is $16$ where the inter-group interference is more effective. When there are closely spaced users, the interfering signal needs to be learned and suppressed adaptively (by using the secondary temporal cells) as in the case of AMF, in order to detect MPCs of intended users with high probability. Contrary to AMF, MF does not take the effect of interfering signals to desired MPC, which results in considerable performance degradation. In Fig. \ref{fig:diff_pd}, when $\bar{\beta}^{(g)}$'s are different among different groups (there is $15$ dB difference between the average received power of the users in the weakest and strongest group, i.e., $10\log\left(\frac{\bar{\beta}^{(4)}}{\bar{\beta}^{(1)}}\right)=15$ dB) and $K=8$, it can be seen that the AMF based CFAR thresholding is very robust against the near-far effect. That is to say, when compared to Fig. \ref{fig:same_pd}, average $P_D$ of the weakest users in Group-1 seems to be not effected with the strong interfering signals of other users. However, in MF case, the average $P_D$ of weakest users is significantly reduced such that it cannot exceed $0.3$ even for large values of SNR. This means that the activity of users in Group-1 is missed by MF based thresholding with high probability in most of the time when strong interfering groups are present.\\
\indent In Fig. \ref{fig:s10_same_mse} and Fig. \ref{fig:s11_same_mse}, nMSE performances of different instantaneous CSI estimators, namely RR-MMSE in \eqref{eqn:W_mmse} and BA-LS in \eqref{eqn:W_bals}, for the effective channel are depicted for $K=8$ and $K=16$ cases where $\bar{\beta}^{(g)}$'s are assumed to be equal. It is observed that when AMF based CFAR thresholding is used to obtain the sparsity map and JADPP, the performances of RR-MMSE and BA-LS estimators are very close to each other, and the gap between their performances and that of the RR-MMSE estimator with true CCM is very small. We can conclude that AMF based CFAR thresholding is very effective to construct the sparsity map and CCMs even with use of small amount of slow-time training data. It is seen that the BA-LS with AMF, utilizing the estimated sparsity map only, appears to be so effective such that the benchmark performance is attained (without necessitating the true knowledge of CCMs) at significantly reduced dimensions. On the other hand, MF based thresholding seems to yield performance losses especially for the case of large number of active users. Since MF type JADPP estimation does not take the interfering users into account, it is not possible to locate MPCs on joint angle-delay map accurately enough especially when the slow-time training amount is limited. In this case, since the CCMs are inaccurately constructed, the updated analog beamformer ($\mathbf{S}^{(g)}$) is not effective anymore while suppressing the inter-group interference. This results in significant performance degradation for RR-MMSE and BA-LS estimators based on MF when compared to AMF based thresholding. Here, the superiority of AMF against MF shown by average $P_D$ vs SNR curves in Fig. \ref{fig:pd}, is also validated in terms of channel estimation accuracy.\\
\indent Similarly, in Fig. \ref{fig:diff_mse}, the case of unequal average received power among different groups is investigated, where $10\log\left(\frac{\bar{\beta}^{(4)}}{\bar{\beta}^{(1)}}\right)=15$ dB is taken. In that case, the performance of reduced rank estimators based on AMF still attains that of the true CCM case. The figure confirms the robustness of AMF type thresholding against increased level of inter-group interference. It is still possible to construct CCMs accurately enough irrespective of differences between the received power levels of different groups. However, the performance of MF based estimators seems to be highly sensitive to the unequal power levels among different groups, i.e., the near-far effect is more apparently observed. In this case, MF fails to locate the MPCs for weak users accurately on joint angle-delay map, which leads to poor channel estimation accuracy for them as shown in Fig. \ref{fig:s10_diff_mse_weak}. For strong user groups, the CCMs can be constructed accurately enough so that the performance of MF based channel estimators is close to that of AMF based ones. It can be noted that the findings in Fig. \ref{fig:diff_mse} are highly consistent with the ones given by average $P_D$ curves in Fig. \ref{fig:diff_pd}.\\
\indent In Fig. \ref{fig:s10_same_T_30dB}, nMSE performances for different estimators are given as a function of $T_{fast}$ when the length of slow-time training data $T$ used for the acquisition of CCMs and sparsity map is fixed to $32$. Here, it is assumed that due to the mobility of users, the instantaneous CSI may change more rapidly compared to their JADPPs. Thus, in fast-time channel acquisition mode, small amount of training data is sufficient to estimate instantaneous CSI by exploiting previously acquired slowly varying parameters. When the sparsity map is obtained accurately enough with the use of AMF based CFAR thresholding, RR-MMSE and BA-LS, which are aware of the spatial signatures of the channels, necessitates significantly reduced amount of fast-time training when compared to the conventional LS type estimator, which is unaware of the jointly sparse structure of the MIMO channel. On the other hand, conventional LS type estimator shows superior performance to BA-LS type for larger fast-time training data when MF based thresholding is adopted. That is because of the inaccurate construction of the sparsity map, which degrades the performance of BA-LS type estimator significantly in case of MF based thresholding.\\
\indent In Fig. \ref{fig:s10_same_snapshot}, nMSE performances of different estimators are shown against different values of $J$ which is used to construct the JADPP before two-stage CFAR thresholding in \eqref{eq:J} when there is $8$ active users with equal average received power. Here, $T=T_{fast}=32$ are taken. As it can be seen, a very small amount of training snapshots is sufficient for AMF based estimators in order to sustain nearly optimal performance (given by RR-MMSE with true CCM). In contrast, MF based estimators require much larger amount of slow-time training snapshots to guarantee accurate enough CCM construction. Therefore, the considerable advantage of the proposed AMF based reduced rank estimators in terms of training overhead over the conventional ones is highlighted both for fast-and-slow time training modes.

\section{Conclusions}
\label{sec:conc}

In this paper, we proposed a novel JADPP and sparsity map estimator based on which CCMs of MPCs for each user are constructed in full dimension first, and then effective beam and channel acquisition are carried out for SC wideband massive MIMO systems employing hybrid beamforming architecture. We contribute the literature in many aspects. First, a novel adaptive spatio-temporal matched filtering type estimator and CFAR thresholding, inspiring from the radar literature, was provided to construct JADPPs and sparsity map of massive MIMO channel in SC transmission. Then, by using significantly reduced amount of training data, fully parametric construction of CCMs was fulfilled based on the estimated sparsity map and power intensities on joint angle-delay plane. Second, a nearly optimal statistical pre-beamformer design, exploiting estimated CCMs, was proposed in order to reduce the channel dimension effectively for JSDM framework. Following the pre-beamforming stage, two different subspace-aware instantaneous CSI estimators, whose training overhead is independent of the number of users in the system were designed. It was shown that the proposed estimators attain the benchmark performance of the MMSE estimator with true knowledge of covariance matrices.

\section*{Appendix I: Derivation of AMF in \eqref{eq:amf_est}}
\label{sec:Appendix_1}
Let the $i^{th}$ column of $\tilde{\mathbf{N}}_l^{(k)}$ in \eqref{eq:YNtilde} be $\bm{\eta}_{l,i}^{(k)}\triangleq\left[\tilde{\mathbf{N}}_l^{(k)}\right]_{(:,i)}$. By using (\ref{eqn:mimo_multipath_correlations}) and assuming that random training sequences are used (pseudo-random code), i.e., $\mathbb{E}\left\{x_n^{(k)}\left(x_{n'}^{(k')}\right)^*\right\}=\delta_{kk'}\delta_{nn'}$, we can write the correlation matrix of the received sequence after hybrid beamforming $\tilde{\mathbf{y}}_n=\mathbf{U}^H\mathbf{y}_n$ in \eqref{eqn:multiuser_multipath_mimo_model} as
\begin{equation}
\mathbf{R}_{\tilde{y}}\triangleq\mathbb{E}\left\{\tilde{\mathbf{y}}_n\tilde{\mathbf{y}}_n^H\right\}=\sum_{k=1}^{K}\sum_{l=0}^{L-1}\mathbf{U}^H\mathbf{R}_l^{(k)}\mathbf{U}+N_o\mathbf{I}_D.
\end{equation}
Then, it can be shown that $\mathbb{E}\left\{\bm{\eta}_{l,m}^{(k)}\left(\bm{\eta}_{l,n}^{(k)}\right)^H\right\}=\mathbf{R}_{\bm{\eta},l}^{(k)}\delta_{mn}$ where $\mathbf{R}_{\bm{\eta},l}^{(k)}=\mathbf{R}_{\tilde{y}}-\mathbf{U}^H\mathbf{R}_l^{(k)}\mathbf{U}$. Here, $\mathbf{R}_{\bm{\eta},l}^{(k)}$ can be regarded as the spatial autocorrelation matrix of interfering MPCs other than the one located in CUT. By using \eqref{eq:Y} and $\mathbf{R}_{\bm{\eta},l}^{(k)}$, we can write the conditional pdf of $\tilde{\mathbf{Y}}$ in \eqref{eq:YNtilde} as follows
\begin{equation}
\label{eq:multivariate_eqn}
p\left(\mathbf{\tilde{Y}}\mathrel{\Big|}\mathbf{R}_{\bm{\eta},l}^{(k)},\alpha_l^{(k)}\right)=
\frac{\exp\left(\scalebox{0.8}[1.0]{\( - \)}\operatorname{Tr}\left\{\left[\mathbf{R}_{\bm{\eta},l}^{(k)}\right]^{\!{-1}}\left(\tilde{\mathbf{Y}}\scalebox{0.8}[1.0]{\( - \)}\alpha_l^{(k)}\tilde{\mathbf{u}}(\phi_i)(\mathbf{x}_l^{(k)})^{\!{H}}\right)\!\left(\tilde{\mathbf{Y}}\scalebox{0.8}[1.0]{\( - \)}\alpha_l^{(k)}\tilde{\mathbf{u}}(\phi_i)(\mathbf{x}_l^{(k)})^{\!{H}}\right)^H\right\}\right)}{\pi^{NT}\left[\det\left(\mathbf{R}_{\bm{\eta},l}^{(k)}\right)\right]^T}\cdot
\end{equation}
\noindent The ML estimate of $\alpha_l^{(k)}$ can be obtained as a result of the minimization problem given below
\begin{align}
&\hat{\alpha}_l^{(k)}=\operatorname*{arg\,max}_{\alpha_l^{(k)}}\ \!\operatorname*{max}_{\mathbf{R}_{\bm{\eta},l}^{(k)}}\ln{p\left(\mathbf{\tilde{Y}}\mathrel{\Big|}\mathbf{R}_{\bm{\eta},l}^{(k)},\alpha_l^{(k)}\right)} \label{eq:min_min} \\ 
\!&=\!\operatorname*{arg\,min}_{\alpha_l^{(k)}}\>\operatorname*{min}_{\mathbf{R}_{\bm{\eta},l}^{(k)}}\left\{T\ln\det\left[\mathbf{R}_{\bm{\eta},l}^{(k)}\right]\!+\!\operatorname{Tr}\left\{\left[\mathbf{R}_{\bm{\eta},l}^{(k)}\right]^{\!{-1}}\!\left(\tilde{\mathbf{Y}}\!-\!\alpha_l^{(k)}\tilde{\mathbf{u}}(\phi_i)(\mathbf{x}_l^{(k)})^H\right)\!\!\left(\tilde{\mathbf{Y}}\!-\!\alpha_l^{(k)}\tilde{\mathbf{u}}(\phi_i)(\mathbf{x}_l^{(k)})^H\right)^H\right\}\!\right\}.\nonumber
\end{align}
In order to solve \eqref{eq:min_min}, the following matrix identities can be used: for given arbitrary matrices $\mathbf{A}$ and $\mathbf{X}$, we can write $\frac{\partial}{\partial \mathbf{X}}\ln\left(\det\left(\mathbf{X}\right)\right) = \mathbf{X}^{-1}$ and $\frac{\partial}{\partial \mathbf{X}}\operatorname{Tr}\left\{\mathbf{X}^{-1}\mathbf{A}\right\} = -\left(\mathbf{X}^{-1}\mathbf{A}\mathbf{X}^{-1}\right)^T$ \cite{petersen07}. By taking the partial derivative of the cost function in \eqref{eq:min_min} with respect to $\mathbf{R}_{\bm{\eta},l}^{(k)}$ and equating it to $\mathbf{0}$, one get
\begin{equation}
\hat{\mathbf{R}}_{\bm{\eta},l}^{(k)} = \frac{\left(\tilde{\mathbf{Y}}\ \scalebox{0.8}[1.0]{\( - \)}\ \alpha_l^{(k)}\tilde{\mathbf{u}}(\phi_i)(\mathbf{x}_l^{(k)})^H\right)\left(\tilde{\mathbf{Y}}\ \scalebox{0.8}[1.0]{\( - \)}\ \alpha_l^{(k)}\tilde{\mathbf{u}}(\phi_i)(\mathbf{x}_l^{(k)})^H\right)^H}{T}
\label{eq:min_Reta}
\end{equation}
\noindent which is simply the ML estimate of the autocorrelation matrix by using $T$ samples. Then, by replacing $\mathbf{R}_{\bm{\eta},l}^{(k)}$ in \eqref{eq:min_min} with $\hat{\mathbf{R}}_{\bm{\eta},l}^{(k)}$ in \eqref{eq:min_Reta}, the minimization in \eqref{eq:min_min} can be simplified as
\begin{equation}
\hat{\alpha}_l^{(k)} =\operatorname*{arg\,min}_{\alpha_l^{(k)}}\>\det\left[\left(\tilde{\mathbf{Y}}-\alpha_l^{(k)}\tilde{\mathbf{u}}(\phi_i)(\mathbf{x}_l^{(k)})^H\right)\left(\tilde{\mathbf{Y}}-\alpha_l^{(k)}\tilde{\mathbf{u}}(\phi_i)(\mathbf{x}_l^{(k)})^H\right)^H\right].
\label{eq:min_alpha}
\end{equation}
To simplify the above minimization, the inner term can be expressed in the following quadratic form
\begin{equation}
\hat{\alpha}_l^{(k)} =\operatorname*{arg\,min}_{\alpha_l^{(k)}}\>\det\left[\left(\mathbf{z}-\mathbf{b}\alpha_l^{(k)}\right)\left(\mathbf{z}-\mathbf{b}\alpha_l^{(k)}\right)^H+\mathbf{\Psi}\right]
\label{eq:min_alpha_quad}
\end{equation}
where $\mathbf{z} = \tilde{\mathbf{Y}}\mathbf{x}_l^{(k)}{\big/}\left\Vert\mathbf{x}_l^{(k)}\right\Vert^2$, $\mathbf{b} = \tilde{\mathbf{u}}(\phi_i)\left\Vert\mathbf{x}_l^{(k)}\right\Vert^2$ and $\mathbf{\Psi}$ is given in \eqref{eq:amf_est}. Then, we can modify \eqref{eq:min_alpha_quad} as
\begin{align}
\hat{\alpha}_l^{(k)} &=\operatorname*{arg\,min}_{\alpha_l^{(k)}}\>\det\left[\mathbf{\Psi}^{1/2}\left[\mathbf{\Psi}^{-1/2}\left(\mathbf{z}-\mathbf{b}\alpha_l^{(k)}\right)\left(\mathbf{z}-\mathbf{b}\alpha_l^{(k)}\right)^H\mathbf{\Psi}^{-1/2}+\mathbf{I}\right]\mathbf{\Psi}^{1/2}\right] \nonumber \\
&=\operatorname*{arg\,min}_{\alpha_l^{(k)}}\>\det\left[\mathbf{\Psi}^{-1/2}\left(\mathbf{z}-\mathbf{b}\alpha_l^{(k)}\right)\left(\mathbf{z}-\mathbf{b}\alpha_l^{(k)}\right)^H\mathbf{\Psi}^{-1/2}+\mathbf{I}\right].
\label{eq:alpha_1}
\end{align}
By using the Sylvester's Theorem in \cite{petersen07}, where $\det(\mathbf{I}+\mathbf{AB})=\det(\mathbf{I}+\mathbf{BA})$, \eqref{eq:alpha_1} can be rearranged as
\begin{align}
\hat{\alpha}_l^{(k)} &=\operatorname*{arg\,min}_{\alpha_l^{(k)}}\>\det\left[1+\left(\mathbf{z}-\mathbf{b}\alpha_l^{(k)}\right)^H\mathbf{\Psi}^{-1}\left(\mathbf{z}-\mathbf{b}\alpha_l^{(k)}\right)\right] \nonumber \\
&=\operatorname*{arg\,min}_{\alpha_l^{(k)}}\>\left(\mathbf{z}-\mathbf{b}\alpha_l^{(k)}\right)^H\mathbf{\Psi}^{-1}\left(\mathbf{z}-\mathbf{b}\alpha_l^{(k)}\right)=\frac{\mathbf{b}^H\mathbf{\Psi}^{-1}\mathbf{z}}{\mathbf{b}^H\mathbf{\Psi}^{-1}\mathbf{b}}.
\end{align}
Finally, the ML estimate of the complex channel gain of $l^{th}$ MPC of $k^{th}$ user can be obtained as in the compact form given in \eqref{eq:amf_est}.

\ifCLASSOPTIONcaptionsoff
  \newpage
\fi

\bibliographystyle{ieeetr}
\bibliography{IEEE_TCOM}

\end{document}